\documentclass[useAMS,usenatbib]{mn2e}

\usepackage{amstext}
\usepackage{amsmath, float}
\usepackage[utf8x]{inputenc}
\usepackage{graphicx}
\usepackage{bm}
\usepackage{booktabs}
\usepackage{comment}

\newcommand{\be}{\begin{equation}}
\newcommand{\ee}{\end{equation}}
\newcommand{\bdm}{\begin{displaymath}}
\newcommand{\edm}{\end{displaymath}}
\newcommand{\bea}{\begin{eqnarray}}
\newcommand{\eea}{\end{eqnarray}}
\newcommand{\ba}{\begin{align}}
\newcommand{\ea}{\end{align}}

\title[GRMHD dynamo in thick accretion disks]
{General relativistic magnetohydrodynamic dynamo in thick accretion disks: fully nonlinear simulations}
\author[N. Tomei, L. Del Zanna, M. Bugli, N. Bucciantini]
{N. Tomei$^{1,2,3}$\thanks{E-mail: tomei@arcetri.astro.it}, 
L. Del Zanna$^{1,2,3}$,
M. Bugli$^{4}$,
N. Bucciantini$^{2,1,3}$\\
$^{1}$Dipartimento di Fisica e Astronomia, Universit\`a di Firenze, 
Via G. Sansone 1, 50019 Sesto Fiorentino (Firenze), Italy\\
$^{2}$INAF - Osservatorio Astrofisico di Arcetri, Largo E. Fermi 5, 50125 Firenze, Italy\\
$^{3}$INFN - Sezione di Firenze, Via G. Sansone 1, 50019 Sesto Fiorentino (Firenze), Italy\\
$^{4}$Département d’Astrophysique, CEA - Saclay, Orme des Merisiers, F-91191 Gif-sur-Yvette, France}

\begin{document}


\pagerange{\pageref{firstpage}--\pageref{lastpage}} \pubyear{2019}

\maketitle

\label{firstpage}

\begin{abstract}
The recent imaging of the M87 black hole at millimeter wavelengths by the \emph{Event Horizon Telescope} (EHT) collaboration has triggered a renewed interest in numerical models for the accretion of magnetized plasma in the regime of general relativistic magnetohydrodynamics (GRMHD). Here \emph{non-ideal} simulations, including both the resistive effects and, above all, the \emph{mean-field dynamo} action due to sub-scale, unresolved turbulence, are applied for the first time to such systems in the fully nonlinear regime. Combined with the differential rotation of the disk, the dynamo process is able to produce an exponential growth of any initial seed magnetic field up to the values required to explain the observations, when the instability tends to saturate even in the absence of artificial quenching effects. Before reaching the final saturation stage we observe a secondary regime of exponential growing, where the magnetic field increases more slowly due to accretion, which is modifying the underlying equilibrium. By varying the dynamo coefficient we obtain different growth rates, though the field seems to saturate at approximately the same level, at least for the limited range of parameters explored here, providing substantial values for the \emph{MAD parameter} for magnetized accretion. For reasonable values of the central mass density and the commonly employed recipes for synchrotron emission by relativistically hot electrons, our model is able to reproduce naturally  the observed flux of Sgr~A*, the next target for EHT.
\end{abstract}

\begin{keywords}
magnetic fields -- (\emph{magnetohydrodynamics}) MHD -- dynamo -- relativistic processes -- 
black hole physics -- accretion, accretion disks.
\end{keywords}
%

\section{Introduction}

Large-scale magnetic fields are well known to play a fundamental role in high-energy astrophysics, hence a natural question one needs to answer is how fields in sources such as compact stars or accretion disks are amplified from initial seed values. Battery-like mechanisms are needed to create primordial extragalactic fields \citep{Kronberg:1994}, which may be amplified to higher values by plasma advection, rotation and collapse to values appropriate for stellar magnetism \citep{Mestel:1999}, up to $B\sim 10^{12}$~G, the field of a \emph{standard} neutron star, a value required to power the surrounding young supernova remnant, when present, as recognized even before the actual discovery of pulsars \citep{Pacini:1967}. 

In most cases a further amplification may occur by instabilities capable of converting kinetic energy into magnetic energy. Under certain conditions, within the ideal magnetohydrodynamical regime (MHD), both Kelvin-Helmoltz and Rayleigh-Taylor instabilities are able to provide such effect, and these have been studied also in the relativistic environment of Pulsar Wind Nebulae \citep{Bucciantini:2004,Bucciantini:2006a}. A very efficient and ideal process, operating in differentially rotating systems when seed fields are already present, is the  \emph{magnetorotational instability} \citep[MRI,][]{Balbus:1998}, known to be able to amplify the fields on ideal timescales and to trigger MHD turbulence.

However, the main process believed to be responsible for magnetic field amplification is a \emph{non-ideal} one, thus requiring a modification of the MHD equations. In typical astrophysical plasmas, this effect is due to the nonlinear coupling of small-scale velocity and magnetic field fluctuations, possibly caused by the instabilities mentioned above. The result of this correlation leads to the creation of an electromotive force capable of amplifying magnetic fields. This process is known as \textit{mean-field dynamo} \citep[e.g.][]{Moffatt:1978,Krause:1980,Cowling:1981,Roberts:1992,Brandenburg:2005} and has been applied to a large number of astrophysical contexts, from the Sun \citep{Parker:1955} to cosmological fields \citep{Kulsrud:1997}. 

Descending into deeper details, the new electromotive term due to the correlated turbulent fluctuations, to be plugged into the MHD induction equation for the magnetic field evolution, can be written as 
\be
<\delta \bm{v} \times \delta\bm{B}> = \alpha_\mathrm{dyn}\bm{B} - \beta_\mathrm{dyn} \bm{J},
\ee
where the two coefficients would be tensors in the most general case, but are usually assumed to be scalars for simplicity \citep[e.g.][]{Krause:1980}. In axisymmetric, differentially rotating systems, the first term describes the so-called $\alpha$-\textit{effect}, responsible for the creation of poloidal fields starting from toroidal fields (prohibited in any \emph{ideal} MHD axisymmetric configuration); the second one is an additional resistive term, enhancing the dissipation of the current density $\bm{J}=\bm{\nabla}\times\bm{B}$. The combined action allows one to close the dynamo cycle, that supports the amplification of magnetic fields according to the mechanism known as $\alpha-\Omega$ dynamo. The same result can be achieved by employing a \emph{non-ideal}, modified Ohm's law with a novel term proportional to the mean magnetic field, that is
\be
\label{eq:ohm-classic}
\bm{E}' = \xi \bm{B} + \eta\bm{J},
\ee
where $\bm{E}' = \bm{E} + \bm{v}\times\bf{B}$ is the comoving electric field, $\xi = - \alpha_\mathrm{dyn}$ is the proper dynamo action term, and $\eta$ the combined Ohmic and turbulent resistivity coefficient.

As far as compact objects are concerned, the $\alpha-\Omega$ dynamo may be responsible for the amplification of fields up to $B\sim 10^{14-15}$~G in magnetars during the proto-neutron star phase. If the initial rotation period is less than 1~ms, the field can be amplified either by differential rotation, magnetic instabilities, and the dynamo effects described above \citep{Duncan:1992,Bonanno:2003,Burrows:2007,Obergaulinger:2014,Guilet:2015,Mosta:2015}. The intense magnetic fields that form can be so strong to cause quadrupolar deformations in the star, that in some cases are able to radiate gravitational waves \citep{Dall'Osso:2009,Pili:2017} and guide relativistic outflows capable of powering gamma-ray bursts \citep{Usov:1992,Bucciantini:2009,Metzger:2011,Bucciantini:2012,Pili:2016}. 

Certainly the amplification of the magnetic field plays a key role in the accretion around the black holes of AGNs, \emph{Active Galactic Nuclei} \citep[e.g.][]{Pariev:2007}. From a theoretical point of view it is believed that disks are threaded by magnetic fields allowing MRI to operate, leading to a redistribution of angular momentum and to MHD turbulence. MRI works for both ordered and disordered fields, though this mechanism alone is inefficient in compensating the turbulent decay if the initial fields are too weak or too incoherent \citep{Bhat:2017}. In these cases a genuine dynamo process would be actually needed to amplify the initial fields to higher values.

The MRI-induced turbulence triggers in turn the accretion of mass and magnetic flux towards the black hole. If the latter is rotating, Poynting-dominated relativistic jets may be launched due to rotational energy extraction inside the ergosphere \citep{Blandford:1977,McKinney:2004}, though even the rotation of the disk itself is known to be able to drive centrifugally driven polar outflows \citep{Blandford:1982}. Amplification of magnetic fields due to dynamo and/or MRI is also important in other scenarios of high-energy astrophysics, like neutron star mergers \citep{Rezzolla:2011,Giacomazzo:2015} and tidal disruption events \citep{Sadowski:2016,Bonnerot:2017}.

The importance of MRI has been recently highlighted in \citet{Bugli:2018}, where the interaction between MRI and the fluid, non-axisymmetric \emph{Papaloizou-Pringle instability} \citep[PPI,][]{Papaloizou:1984} has been investigated. Three-dimensional general relativistic MHD (GRMHD) simulations show that in a magnetized torus around a Schwarzschild black hole, PPI large-scale modes are suppressed by the development of MRI, which is then to be considered the main driver for accretion onto supermassive black holes. 

GRMHD simulations of MRI-induced accretion onto rotating black holes is being receiving considerable attention due to the success of the \emph{Event Horizon Telescope} (EHT) collaboration, capable of imaging the emission and the \emph{shadow} around the event horizon of a black hole for the very first time \citep{EHTCollaboration:2019a}. The comparison between data and numerical models has allowed to infer the main physical quantities of the compact object \citep{EHTCollaboration:2019}. The observed source has been the nucleus of the elliptical galaxy M87, but research is ongoing for the main target Sgr A*, the compact radio source located at the center of our Milky Way, whose emission is believed to be due to (inefficient) accretion onto a black hole of a few $10^6 M_\odot$ \citep{Narayan:1998,Moscibrodzka:2014}. 

A thick torus with a weak magnetic field is typically chosen as initial equilibrium for GRMHD simulations of the accretion onto black holes \citep{DeVilliers:2003,McKinney:2004,McKinney:2006,Hawley:2006}. Two configurations are mainly considered for the initial magnetic field: one leading to SANE, \emph{Standard and Normal Evolution} \citep{Narayan:2012}, or to a MAD one, \emph{Magnetically Arrested Disk} \citep{Narayan:2003}. The latter is characterized by a higher advected magnetic flux with respect to the SANE case, capable of lowering the mass accretion rate down to $\sim 10^{-6}\dot{M}_\mathrm{Edd}$, a factor up to 50 times lower than for SANE accretion  \citep{EHTCollaboration:2019}. The results of the simulations are then post-processed by general relativistic ray-tracing codes solving radiative-transfer in curved spacetimes \citep[e.g.][]{Moscibrodzka:2018}, so that synthetic images can finally be created. 

In this context it is important to underline the capabilities of a GRMHD code in dealing with accretion onto a rotating black hole in Kerr metric, a problem with many numerical difficulties and subtleties (e.g. the use of horizon-penetrating coordinates, the treatment of low-density polar funnel where Poynting jets are launched, and so on). An extensive comparison has been recently pursued among the codes available worldwide, including the one from our group, \texttt{ECHO} \citep{DelZanna:2007}, employed for the simulations in the present paper. The good agreement between the results on a standard test on SANE-type accretion reveals that the community of GRMHD codes is able to provide consistent solutions to address these problems and meaningful comparison with the observations \citep[see][and references therein]{Porth:2019}.

However, it must be noticed that the simulations employed by the EHT community and for the code comparison tests all rely on initial magnetic fields with pressure which is one hundredth of the kinetic pressure. This value corresponds to a subdominant field, but it is not far from the value needed to reproduce both the dynamics needed to launch the polar jets and the non-thermal synchrotron emission (once a model for the distribution for the emitting electrons has been established). A more natural scenario would be the one in which a very low initial field is evolved and amplified by some kind of dynamo process, so to reach \emph{self-consistently} the correct threshold for MRI to induce accretion and to reproduce the correct dynamical and emission properties.

The first relativistic models for mean-field $\alpha-\Omega$ dynamo effects in thick disks around Kerr black holes were proposed more than  two decades ago \citep{Khanna:1996,Brandenburg:1996}, whereas the first implementation within a full GRMHD scheme (even including radiation) is due to \citet{Sadowski:2015}, where the dynamo action in accretion disks was parametrized using the results of local shearing box simulations of MRI, leading to the addition of both poloidal and toroidal magnetic field components. Interestingly, the 2D axisymmetric simulations with the mentioned dynamo recipes were able to produce the same kind of fields obtained in 3D simulations without artificial terms, where the turbulent dynamo is fully operative.

The first rigorous treatment of the dynamo effects within (resistive) GRMHD was presented by \citet{Bucciantini:2013}, later applied to the physics of accretion disks by \citet{Bugli:2014}. There the evolution of magnetic fields was studied by in axisymmetry and in the \emph{kinematic regime}, that is by solving Maxwell equations alone not considering the feedback on the disk's plasma. While  the $\alpha$-effect is supposed to be given by the coupling between unresolved velocity and field fluctuations, the $\Omega$-effect is due to the differential rotation of the disk. The magnetic field threading the disk is indeed amplified exponentially (the growth is unbounded in the kinematic case), not related to the period of rotation (thus on gravity or on the fluid properties), but rather on the microphysics of turbulence, providing the $\xi$ dynamo coefficient in Eq.~(\ref{eq:ohm-classic}). When an odd function of the latitude with respect to the equator is chosen for $\xi$, the model is even capable of reproducing the counterpart of \emph{butterfly diagrams} as observed for the solar cycle \citep{Charbonneau:2010}.

The dynamo in relativistic plasmas is characterized by additional difficulties compared to the classical MHD, since, as for the resistive case, the electric field must be also evolved and \emph{stiff} terms in the equations appear, requiring some sort of implicit treatment \citep[e.g.][]{Palenzuela:2009,Dionysopoulou:2013,DelZanna:2016,Mignone:2019}. The theory and the best numerical strategies are outlined in \citet{Bucciantini:2013}, where a fully covariant generalization of Eq.~(\ref{eq:ohm-classic}) was first proposed for relativistic plasmas, the GRMHD equations with both resistive and dynamo terms were first written in the so-called $3+1$ formalism as needed for numerical integration \citep[see also][]{DelZanna:2018}, and where a robust method for the solution of the implicit step coupled to the inversion procedure from conservative to primitive variables, using a 3-D Newton-Raphson scheme, was suggested.

In this paper, we generalize our previous work on the mean-field dynamo in accreting disks \citep{Bugli:2014} by investigating the completely self-consistent and nonlinear regime (dynamic regime) during the accretion phase. The linear growth of the fields cannot reasonably continue for an arbitrarily long time, and it is expected to be quenched naturally by the feedback on the disk. Our goal is to see how the transition to the non-linear phase occurs, to investigate the interplay with MRI, and the effect of accretion on the dynamo process itself, for a given disk model and a variety of dynamo parameters. 

Finally, on top of our GRMHD model based on the dynamo action, we compute the local emissivity and integrated flux in the radio band (in the approximation of an optically thin plasma), and we propose a comparison with observational data for Sgr A*, in view of the long-awaited analysis by the EHT collaboration.

\section{Equations and numerical methods}

\subsection{The ECHO code for non-ideal GRMHD}

The \textit{Eulerian Conservative High-Order} code, \texttt{ECHO}, is an efficient finite-difference shock-capturing scheme for the GRMHD system of conservation laws, based on the $3+1$ formalism of numerical relativity and working in any spacetime metric \citep{DelZanna:2007}. Here we describe the implementation of the non-ideal effects, namely the inclusion of the resistive and dynamo terms, following \citet{Bucciantini:2013, DelZanna:2018}.

In the $3+1$ formalism, any four-dimensional spacetime must be split according to the metric
\be
ds^2=-\alpha^2dt^2+\gamma_{ij}(dx^i+\beta^idt)(dx^j+\beta^jdt),
\ee
where $\alpha$ is the \textit{lapse function}, $\beta^i$ the \textit{shift vector} and $\gamma_{ij}$ is the 3-metric, used to raise/lower the indexes of any spatial three-dimensional vector or tensor. Within this formalism, the system of dynamo-resistive GRMHD equations in conservative form is
\be
\begin{split}
 \partial_t  & (\sqrt{\gamma} D)  +  \partial_k [ \sqrt{\gamma} ( \alpha v^k - \beta^k ) D ]  = 0,  \\
 \partial_t & (\sqrt{\gamma} S_i)  +  \partial_k [ \sqrt{\gamma} ( \alpha S^k_{\,i} - \beta^kS_i)]  \\
& = \sqrt{\gamma} [ \tfrac{1}{2}\alpha S^{lm}\partial_i\gamma_{lm} + 
S_k\partial_i\beta^k - (\mathcal{E} + D)\partial_i\alpha ],  \\
\partial_t & (\sqrt{\gamma} \mathcal{E}) +  
\partial_k [ \sqrt{\gamma} ( \alpha S^k - \beta^k \mathcal{E} - \alpha v^k D )]    \\
& = \sqrt{\gamma} [ \tfrac{1}{2}S^{lm}(\beta^k\partial_k\gamma_{lm} - \partial_t\gamma_{lm}) + S^l_{\,m}\partial_l\beta^m - S^k \partial_k\alpha ] ,  \\
\partial_t & (\sqrt{\gamma} E^i)  - \sqrt{\gamma} \, \epsilon^{ijk}  \partial_j ( \alpha B_k - \epsilon_{klm} \beta^l E^m) +\sqrt{\gamma}  q  ( \alpha v^i - \beta^i )  \\
& = - \sqrt{\gamma} \alpha\Gamma  \, \eta^{-1}  [ \, E^i+\epsilon^{ijk}v_jB_k- E^jv_j\, v^i  \\
& - \xi ( B^i-\epsilon^{ijk}v_jE_k- B^jv_j\, v^i ) \, ] ,  \\
\partial_t & (\sqrt{\gamma} B^i)  +  \sqrt{\gamma} \, \epsilon^{ijk}  \partial_j ( \alpha E_k + \epsilon_{klm} \beta^l B^m) = 0, 
\label{eq:grmhd}
\end{split}
\ee
with the additional non-evolutionary constraints
\be
\partial_k (\sqrt{\gamma} E^k ) = \sqrt{\gamma} q, \quad
\partial_k (\sqrt{\gamma} B^k ) =  0.
\ee
In the above expressions, $D=\rho\Gamma$ is the mass density in the laboratory frame, $S_i=w \Gamma^2v_i +\epsilon_{ijk}E^jB^k$ the momentum density, $S_{ij}=w\Gamma^2 v_iv_j+p\gamma_{ij}-E_iE_j-B_iB_j+u_\mathrm{em}\gamma_{ij}$ the stress tensor, $\mathcal{E} + D =w\Gamma^2 - p + u_\mathrm{em}$ the total energy density, $\rho$ the mass density in the comoving frame, $w$ the enthalpy per unit volume, $p$ the thermal pressure, $v^i$ the fluid 3-velocity and $\Gamma$ its corresponding Lorentz factor, $u_\mathrm{em} = \tfrac{1}{2}(E^2+B^2)$ the electromagnetic energy density, $E^i$ and $B^i$ the electric and magnetic fields, $q$ the charge density, whereas $\gamma$ and $\epsilon_{ijk}$ are respectively the determinant and the Levi-Civita pseudo-tensor of the 3-metric. Notice that in the Cowling approximation employed in the remainder of this work, that is for a non-evolving metric, $\sqrt{\gamma}$ may be extracted from all time derivatives and the term $-\tfrac{1}{2}S^{lm}\partial_t\gamma_{lm}$ vanishes in the equation for $\mathcal{E}$. In order to close the system, an equation of state must be also specified. Here we assume the equation of state for a perfect fluid, $p=(\hat{\gamma}-1)\rho\epsilon$, where $\epsilon$ is the thermal energy per unit mass, or equivalently
\be
\label{eq:entalpy}
w = \rho + \frac{\hat{\gamma}}{\hat{\gamma} - 1} \, p = \rho + \hat{\gamma}_1 \, p,
\ee
and $\hat{\gamma}$ is the adiabatic index ($\hat{\gamma}=4/3$ and $\hat{\gamma}_1=4$ for a relativistic fluid).

The evolution equation for the electric field is the one that takes into account the dynamo and resistive effects, through the coefficients $\xi$ and $\eta$. This can be made to descend from the natural and fully covariant generalization of Eq.~\eqref{eq:ohm-classic} \citep[see][]{Bucciantini:2013,DelZanna:2018}:
\be
e^\mu=\xi b^\mu+\eta j^\mu,
\ee
where the 4-vectors $e^\mu$, $b^\mu$, and $j^\mu$ are, respectively, the electric field, magnetic field, and current density in the frame comoving with the fluid. Starting from this expression one derives the new Ohm's law expressed in $3+1$ form as
\be
\begin{split}
&\Gamma (E^i+\epsilon^{ijk}v_jB_k- E^jv_j \,v^i)= \\
& \eta(J^i-qv^i)+\xi\Gamma (B^i-\epsilon^{ijk}v_jE_k-B^jv_j\,v^i),
\end{split}
\ee 
where $J^i$ is the usual conduction current, which has been plugged into the Ampère-Maxwell equation for $E^i$  in the system (\ref{eq:grmhd}). When $\xi = 0$ one recovers the purely resistive version, while the ideal MHD limit for $\eta \to 0$ is obtained by neglecting this equation and by replacing $E^i = - \epsilon^{ijk}v_jB_k$ in fluxes.

The \texttt{ECHO} code is designed to solve numerically the system of Eqs.~(\ref{eq:grmhd}), for any technical detail the reader is referred to \citet{DelZanna:2007}. The code is based on high-order spatial reconstruction and derivation algorithms and a simple two-wave solver. In particular, in the present work we employ the MP5 algorithm, \textit{Monotonicity Preserving} \citep{Suresh:1997}, to reconstruct variables at cell interfaces, combined to a sixth-order fixed-stencil derivation routine for numerical fluxes, allowing to reach an overall fifth order of spatial accuracy in smooth regions. The solenoidal constraint for the magnetic field is enforced through the UCT (\textit{Upwind Constrained Transport}) method based on a staggered representation of magnetic field components \citep{Londrillo:2000,Londrillo:2004}, allowing the preservation of the solenoidal condition to machine accuracy for a second order scheme, or up to its nominal spatial accuracy when higher order methods are employed. The \texttt{ECHO} code for GRMHD has been also extended to  dynamical spacetimes in the asymptotically flat approximation \citep{Bucciantini:2011}.

\subsection{The implicit step and the inversion from conservative to primitive variables}

Here we focus on the implementation of the IMEX (\textit{IMplicit EXplicit}) Runge-Kutta scheme to perform the temporal evolution of the equations, needed to properly treat the \emph{stiff} terms in the equation for the electric field, proportional to $\eta^{-1}\gg1$ \citep{Palenzuela:2009}, regardless of the presence of dynamo action. Let us rewrite the resistive GRMHD equations in the form 
\be
\label{cons_syst}
\partial_t\bm{\mathcal{U}} = - \partial_k\bm{\mathcal{F}}^k + \bm{\mathcal{S}}
= \bm{\mathcal{Q}} + \bm{\mathcal{R}},
\ee
where (we recall that the metric can be time-dependent)
\be
\bm{\mathcal{U}} = \sqrt{\gamma}\, [D,S_i,\mathcal{E},E^i,B^i]^\mathrm{T},
\label{eq:conservate(j)}
\ee
are the \emph{conserved} variables, $\bm{\mathcal{F}}$ the corresponding fluxes, $\bm{\mathcal{S}}$ the source terms, and where we have then merged all non-stiff terms in \bm{\mathcal{Q}} (including flux derivatives), while $\bm{\mathcal{R}}$ contains only terms proportional to $\eta^{-1}\gg1$ , precisely those requiring an implicit treatment. We now split the variables into two subsets, $\bm{\mathcal{U}}=\{\bm{\mathcal{X}}, \bm{\mathcal{Y}}\}$, where $\bm{\mathcal{X}}= \sqrt{\gamma}\bm{E}$ (whose source terms are stiff) and $\bm{\mathcal{Y}}$ refers to the remaining ones, with vanishing stiff terms $\bm{\mathcal{R}}_{\bm{\mathcal{Y}}}=0$. The system is then conveniently written as
\be
\partial_t\bm{\mathcal{X}}=\bm{\mathcal{Q}}_{\bm{\mathcal{X}}}[\bm{\mathcal{U}}]+\bm{\mathcal{R}}_{\bm{\mathcal{X}}}[\bm{\mathcal{U}}], \quad
\partial_t\bm{\mathcal{Y}}=\bm{\mathcal{Q}}_{\bm{\mathcal{Y}}}[\bm{\mathcal{U}}].
\ee
Consider now the update number $n$ of the conservative variables, from $\bm{\mathcal{U}}^n$ to $\bm{\mathcal{U}}^{n+1}$, of a time interval $\Delta t$. Let $s$ be the number of IMEX Runge-Kutta substeps, where each step is labeled with $i=1,2,\ldots s$. For any substep $i$, the update is achieved in two distinct phases:
\begin{itemize}
\item First, intermediate values of the conservative variables are obtained as sums of $i-1$ \emph{purely explicit} terms as follows
\be
\bm{\mathcal{X}}_\star^{(i)}=\bm{\mathcal{X}}^n+\Delta t \sum_{j=1}^{i-1} \left[ \tilde{a}_{ij}\bm{\mathcal{Q}}_{\bm{\mathcal{X}}}[\bm{\mathcal{U}}^{(j)}] + {a}_{ij}\bm{\mathcal{R}}_{\bm{\mathcal{X}}}[\bm{\mathcal{U}}^{(j)}] \right],
\ee
and
\be
\bm{\mathcal{Y}}_\star^{(i)}=\bm{\mathcal{Y}}^n+\Delta t \sum_{j=1}^{i-1}\tilde{a}_{ij}\bm{\mathcal{Q}}_{\bm{\mathcal{Y}}}[\bm{\mathcal{U}}^{(j)}],
\ee
where the two (lower triangular) matrices with coefficients $\tilde{a}_{ij}$ and ${a}_{ij}$ have dimensions $s\times s$.
\item Second, variables $\bm{\mathcal{X}}^{(i)}$, those with stiff source terms, undergo an extra \emph{implicit} evolution step for $j=i$, with $\tilde{a}_{ii}=0$ and $a_{ii}\neq 0$ by definition, so that one must solve
\be
\label{eq:implicit_imex}
\bm{\mathcal{X}}^{(i)}=\bm{\mathcal{X}}_\star^{(i)}+a_{ii}\Delta t \, \bm{\mathcal{R}}_{\bm{\mathcal{X}}}[\bm{\mathcal{X}}^{(i)}, \bm{\mathcal{Y}}_\star^{(i)}], \quad \bm{\mathcal{Y}}^{(i)}=\bm{\mathcal{Y}}_\star^{(i)}.
\ee
\end{itemize}
Notice that at the first substep, for $i=1$, only the implicit step is needed. Once all $s$ implicit and $s-1$ explicit contributions are calculated, the final update $\bm{\mathcal{U}}^{n+1}$ is given by
\be
\bm{\mathcal{U}}^{n+1}=\bm{\mathcal{U}}^{n}+\Delta t \sum_{i=1}^{s} \left[ \tilde{\omega}_i\bm{\mathcal{Q}}[\bm{\mathcal{U}}^{(i)}] + {\omega}_i\bm{R}[\bm{\mathcal{U}}^{(i)}] \right],
\ee
where $\tilde{\omega}_i$ and ${\omega}_i$, for $i=1,2,\ldots s$, are additional coefficients required by the IMEX scheme. Here we use the IMEX \emph{Strong Stability Preserving} scheme SSP3(4,3,3) \citep{Pareschi:2005}, with $s=4$ implicit substeps, $\tilde{\omega}_i={\omega}_i$ (and $\tilde{\omega}_1={\omega}_1=0$), and with a third-order accuracy in time \citep[for numerical tests see][]{DelZanna:2014}. 

We now show how the implicit step is carried out in \texttt{ECHO}. From the inversion of the relation \eqref{eq:implicit_imex}, an explicit expression for the electric field as a function of the velocity and known quantities can be derived as shown in \citet{Bucciantini:2013}\footnote{Equation (35) in  \citet{Bucciantini:2013} contains an error in the purely resistive term, fortunately not in the corresponding part of the code. The correct version can also be found in \citet{DelZanna:2016}.}. Here we choose to write it in the form
\be
\label{eq:implicit_final}
\begin{split}
A_0 E^i & 
=  \tilde{\eta} E_\star^i +  A_1 (E_{\star k}\tilde{u}^k) \tilde{u}^i + A_2 \epsilon^{ilm} \tilde{u}_l E_{\star m} \\
& + A_3 B^i + A_4 (B_k \tilde{u}^k) \tilde{u}^i + A_5 \epsilon^{ilm} \tilde{u}_l B_m,
 \end{split}
\ee
where all coefficients are function of the Lorentz factor $\Gamma$ (hence of the velocity) alone. See the Appendix for a derivation of the above equation and for the expressions of the coefficients. It is convenient to work with the new spatial vector $\tilde{u}^i=\Gamma v^i$ (not coincident with the spatial component of the 4-velocity, unless $\beta^i\neq 0$ as in the Minkowski or Schwarzschild metrics), and $\tilde{u}_i=\Gamma v_i$, so that $\Gamma^2 = 1 + \tilde{u}^i \tilde{u}_i$. Above we have also defined $E_\star^k = \mathcal{X}_\star^k/\sqrt{\gamma}$, where $\mathcal{X}_\star^k$ are the electric field components, multiplied by $\sqrt{\gamma}$, computed thanks to the previous explicit substeps $j<i$ of the IMEX scheme, or at the previous timestep when $i=1$. Moreover
\be
\tilde{\eta}=\frac{\eta / \alpha}{ a_{ii}\Delta t},
\ee
with $j=i$ referring to the last, implicit substep. Notice that when $\eta=0$ and $\xi=0$ we recover the ideal GRMHD case.

The relation \eqref{eq:implicit_final} allows one to calculate the electric field as a function of the velocity and of the magnetic field at the end of each sub-step. However, since the numerical scheme evolves in time the conserved variables in Eq.~\eqref{eq:conservate(j)}, primitives variables such as the velocity are not readily available and the implicit step must be nested in the non-linear iterative procedure which recovers primitive variables from the above set of conservative ones. Starting with a straightforward guess for $\tilde{u}^{j\,(0)}$, that is the value corresponding to the set of conserved variables at the previous timestep, we proceed by adopting the following Newton-Raphson scheme:
\begin{itemize}
\item 
at any iteration $(k)$, with a value $\tilde{u}^{j\,(k)}$ for the velocity, work out the electric field given by solving Eq. \eqref{eq:implicit_final};
\item 
evaluate the function
\be
\label{eq:momenta}
f_i ( \tilde{u}^j) = \tilde{w} \gamma_{ij}\tilde{u}^j+\epsilon_{ilm}E^lB^m-S_i,
\ee
where the modified enthalpy is
\be
\tilde{w} = w\Gamma = \frac{\Gamma [\mathcal{E}+D-u_\mathrm{em}]- D /\hat{\gamma}_1}{\Gamma^{2}-1/\hat{\gamma}_1},
\ee
obtained using Eq.~(\ref{eq:entalpy}), is also given in terms of conservative variables and velocity components alone;
\item 
evaluate the Jacobian 
\be
\label{eq:jacobian}
J_{ij} = \frac{\partial f_i}{\partial \tilde{u}^j} = \tilde{w} \gamma_{ij} + 
\tilde{u}_i  \frac{\partial \tilde{w}}{\partial\tilde{u}^j} +
\epsilon_{ilm}\frac{\partial E^l}{\partial\tilde{u}^j}B^m,
\ee
where
\be
\frac{\partial \tilde{w}}{\partial \tilde{u}^j} =
\frac{ 
[\mathcal{E} + D - u_\mathrm{em} ] \tilde{u}_j /\Gamma - 2 \tilde{w} \tilde{u}_j 
 -\Gamma E_i (\partial E^i / \partial\tilde{u}^j )
}{\Gamma^2 - 1/\hat{\gamma}_1},
\ee
and where we have used $\partial\Gamma/\partial\tilde{u}^j = \tilde{u}_j /\Gamma$. The expression for the Jacobian of the electric field is more involved. Recalling that for any function $f(\Gamma)$ we have $\partial f/\partial u^j = \dot{f} \, u_j/\Gamma$, one can write
\begin{eqnarray}
\label{eq:E_jacobian} 
&  A_0 \Gamma (\partial E^i / \partial\tilde{u}^j ) = \\
& - \dot{A}_0 E^i u_j + A_1 \Gamma u^i E_{\star j} 
+ \dot{A}_3 B^i u_j + A_4\Gamma  u^i B_j
\nonumber \\ 
& \! + (A_1 \Gamma \gamma^i_{\,j}+\dot{A}_1 u^i u_j)(E_{\star k}u^k) 
\! + \! \epsilon^{ilm}(A_2\Gamma\gamma_{lj}+\dot{A}_2 u_l u_j) E_{\star m}
\nonumber \\
& \! + (A_4\Gamma \gamma^i_{\,j}+\dot{A}_4 u^i u_j)(B_ku^k)
\! + \! \epsilon^{ilm}(A_5\Gamma \gamma_{lj}+\dot{A}_5 u_l u_j) B_m
\nonumber
\end{eqnarray}
and the expressions for the derivative of the coefficients are reported in the Appendix. Once the full Jacobian is known, we can finally update the velocity as
\be
\tilde{u}^{j \, (k+1)} = \tilde{u}^{j \, (k)} - [J_{ij}^{(k)} ]^{-1} f_i^{(k)} .
\ee
\end{itemize}
The iterations are repeated until the desired accuracy is reached, so that the primitive variables at every substep $j$ can be computed for the IMEX scheme. 

The above 3-D Newton-Raphson scheme based on the vanishing of momentum equations and using the $\tilde{u}_m$ variables, first introduced by \citet{Bucciantini:2013}, has proved to be the most robust one and has been recently adopted in other resistive relativistic MHD codes \citep{Mignone:2019,Ripperda:2019}. The novel feature introduced here is the analytical calculation of the Jacobian in Eq.~(\ref{eq:jacobian}), that appears to be more robust in critical situations and to require $10-20\%$ less iterations compared to the approach in  \citet{Bucciantini:2013}, where the Jacobian was computed numerically, by evaluating the momenta twice for nearby values of $\tilde{u}^j$.

\section{Disk model and numerical set-up}

Our simulations are initialized with the hydrodynamical equilibrium solution for a differentially rotating thick disk, also known as \emph{Polish daughnut} \citep{Kozlowski:1978,Abramowicz:2013}. The general relativistic Euler equation for a rotating fluid in a stationary and axisymmetric gravitational field admits the integral
\label{eq:euler_potential}
\be
W-W_\mathrm{in}+\int\frac{dp}{w}=0,
\ee
where the potential $W$ is defined in terms of the fluid 4-velocity $u^\mu$ as
\be
W=\ln{|u_t|}+\int_0^{\infty}\frac{\Omega d\ell}{1-\ell\Omega},
\ee
where $\ell = - u_\phi/u_t$ is the specific angular momentum, $\Omega = u^\phi/u^t$ the angular velocity, and $W_\mathrm{in}$ the potential evaluated at the inner edge of the disk, $r_\mathrm{in}$. The simplest model is a barotropic disk with constant $\ell$, that is
\be
\label{eq:pressure-disk}
p=Kw^{\hat{\gamma}},~~~~\ell=\ell_0,
\ee
for which
\be
\label{eq:fundamental-disk}
W-W_{in} + K \hat{\gamma}_1 w^{\hat{\gamma}-1} =0,
\ee
with the potential given expressed in terms of the metric as
\be
W = \ln{|u_t|} = \frac{1}{2} \ln\left( \frac{g_{t\phi}^2 - g_{tt} g_{\phi\phi}}{g_{\phi\phi} + 2g_{t\phi}\ell_0 + g_{tt}\ell_0^2}  \right).
\ee
Once the angular moment has been assigned, the potential is defined at each point  in the domain as a function of spherical-like coordinates ($r, \theta$). The matter can fill every closed equipotential surface, that is the surfaces characterized by the condition $W(r,\theta)-W_{in}<0$. Equivalently it is possible to define the potential assigning the position of the cusp, $r_\mathrm{in}$, and of the center of the disk, $r_c$ \citep{DelZanna:2007}, the innermost and outermost points where the angular momentum assumes the Keplerian value. In this way $\ell_0$ is defined by simply evaluating the expression
\be
\ell_0=\pm\frac{r_c^2\mp 2a_\mathrm{BH}\, r^{1/2}_c+a_\mathrm{BH}^2}{r^{3/2}_c-2r_c^{1/2}\pm a_\mathrm{BH}},
\ee
where the upper (lower) sign is for co-rotating (counter-rotating) orbits and $a_\mathrm{BH}$ is the spin parameter of the black hole. Fluid quantities can now be evaluated as
\be
w=w_c\biggl(\frac{W_\mathrm{in}-W}{W_\mathrm{in}-W_c}\biggr)^{\frac{1}{\hat{\gamma}-1}}, \quad
p=p_c\biggl(\frac{w}{w_c}\biggr)^{\hat{\gamma}}, 
\ee
where $p_c = K w_c^{\hat{\gamma}}$, and the density is derived from Eq. (\ref{eq:entalpy}). 

We have considered a disk with hydrodynamical equilibrium so far. In order to study the dynamo amplification mechanism, a small magnetic field must be introduced in such a way that the initial equilibrium is not affected by its presence ($B^2\ll w$). In our simulations a large poloidal loop has been superimposed over the hydrodynamical torus, described by the vector potential \citep{Liska:2018}
\be
A_\phi=A_0\rho^2r^3,
\ee
where $A_0$ is a constant. 

Finally, outside the disk, that is in the region where $W-W_{in}>0$, we introduce a static, unmagnetized and tenuous atmosphere, with density and pressure radial profiles given by \citep{Porth:2017}
\be
\rho_\mathrm{atm}=\rho_\mathrm{min}r^{-3/2}, \quad
p_\mathrm{atm}=p_\mathrm{min}r^{-5/2}.
\ee
The above values are also used to reset density and pressure, respectively, when the numerical inversion from conservative to primitive variables fails in providing physical results.

In Table~\ref{tab:dischi} the parameters used to characterize the disk configuration are shown. Lengths and times are expressed in units of $r_g=GM_\mathrm{BH}/c^2$ (the gravitational radius) and $r_g/c$, respectively. The mass density is normalized against $\rho_0 = n_0 m_p$, where $n_0$ is a reference peak number density in the disk, whereas enthalpy, energy density and fluid and magnetic pressure are normalized against against $\rho_0 c^2$. The spin parameter $a_\mathrm{BH}$ must also be provided in order to characterize the Kerr-type metric, and here we choose the same value used in the aforementioned code comparison project \citep{Porth:2019}.

\begin{table}
\centering
\small
\begin{tabular}{|c|c|c|c|c|c|c|c|}
\hline
$ \boldsymbol{a_\mathrm{BH}}$  & $\boldsymbol{r_\mathrm{in}}$ &  $\boldsymbol{r_{c}}$ & $\boldsymbol{w_{c}}$ & $A_0$ & $\boldsymbol{\rho_\mathrm{min}}$& $\boldsymbol{p_\mathrm{min}}$\\
\hline
0.9375 & 6 & 12 & 1 & $10^{-8}$ & $10^{-4}$ & $10^{-6}$ \\
\hline
\end{tabular}
\vspace{0.1cm}
\caption{Parameters of the initial equilibrium model.}
\label{tab:dischi}
\end{table}

In order to quantify the dynamo action, we can introduce the two characteristic numbers \citep{Bugli:2014}:
\be
C_\Omega=\frac{\Delta\Omega R^2}{\eta},~~~~C_\xi=\frac{\xi R}{\eta},
\ee
where $\Delta\Omega=\Omega_{in}(t=0)-\Omega_c(t=0)$ is a typical angular velocity difference and $R\sim r_c$ is a typical high scale of the torus. These numbers describe the importance of $\alpha$ dynamo and rotation with respect to the dissipation of magnetic fields. The $\eta$ and $\xi$ profiles are chosen so that the diffusion and dynamo processes occur only within the disk. Starting from the maximum values ​​$\xi_\mathrm{max}$ and $\eta_\mathrm{max}$, those actually entering the definition of the dynamo numbers, at each point of the domain we impose
\be
\eta(r,\theta)=\eta_\mathrm{max}S_\eta(r,\theta),
\ee
with 
\be
S_\eta(r,\theta)=\frac{\rho-\rho_\mathrm{atm}}{\rho_\mathrm{max}},
\ee
and
\be
\xi(r,\theta)=
\begin{cases}
\xi_\mathrm{max}S_\xi(r,\theta), & \text{inside the disk,} \\
0 & \text{in the atmosphere,}
\end{cases}
\ee
with
\be
S_\xi(r,\theta)=\frac{\rho\cos\theta}{[\rho\cos\theta]_\mathrm{max}},
\ee
where the presence of an odd function with respect to the equator will lead to a symmetric dynamo action. In Table \ref{tab:run-models} we show the models we have considered for our simulations. We have explored different dynamo numbers $C_\xi$, starting from a reference value (Run1), in order to cover a significant range in the parameter space. In the present analysis we have chosen to leave the hydrodynamical equilibrium, hence the $\Omega (r)$ profile, unchanged as well as a fixed $\eta_\mathrm{max}$, so that $C_\eta$ has also been kept constant.

We adopt here the horizon-penetrating Kerr-Schild metric and 2-D axisymmetric spherical coordinates, \citep[see][for additional details]{Komissarov:2004,Bugli:2018}. The two-dimensional numerical domain extends in the regions delimited by $r_\mathrm{min}=r_h-0.3$, inside the horizon $r_h=1+(1-a^2_\mathrm{BH})^{1/2}$, and $r_\mathrm{max}=100$ in the radial direction, and by $0.06$ and $\pi-0.06$ in the $\theta$ direction. The grid $(512\times 256)$ employed is uniform in $\theta$ but not along the radial direction, where points are defined by the nonlinear function
\be
r_i=r_\mathrm{min}+\frac{r_\mathrm{max}-r_\mathrm{min}}{\Psi}\tan({m_i\arctan{\Psi}}),
\ee
with $m_i$ covering uniformly the range $[0,1]$ and $\Psi$ a \textit{stretching factor} fixed to 10 as in \citet{Bugli:2014}. This choice allows to have a higher resolution in the inner region where larger gradients are expected.

\section{Numerical results}

\begin{figure}
\centering
\includegraphics[width=.5\textwidth]{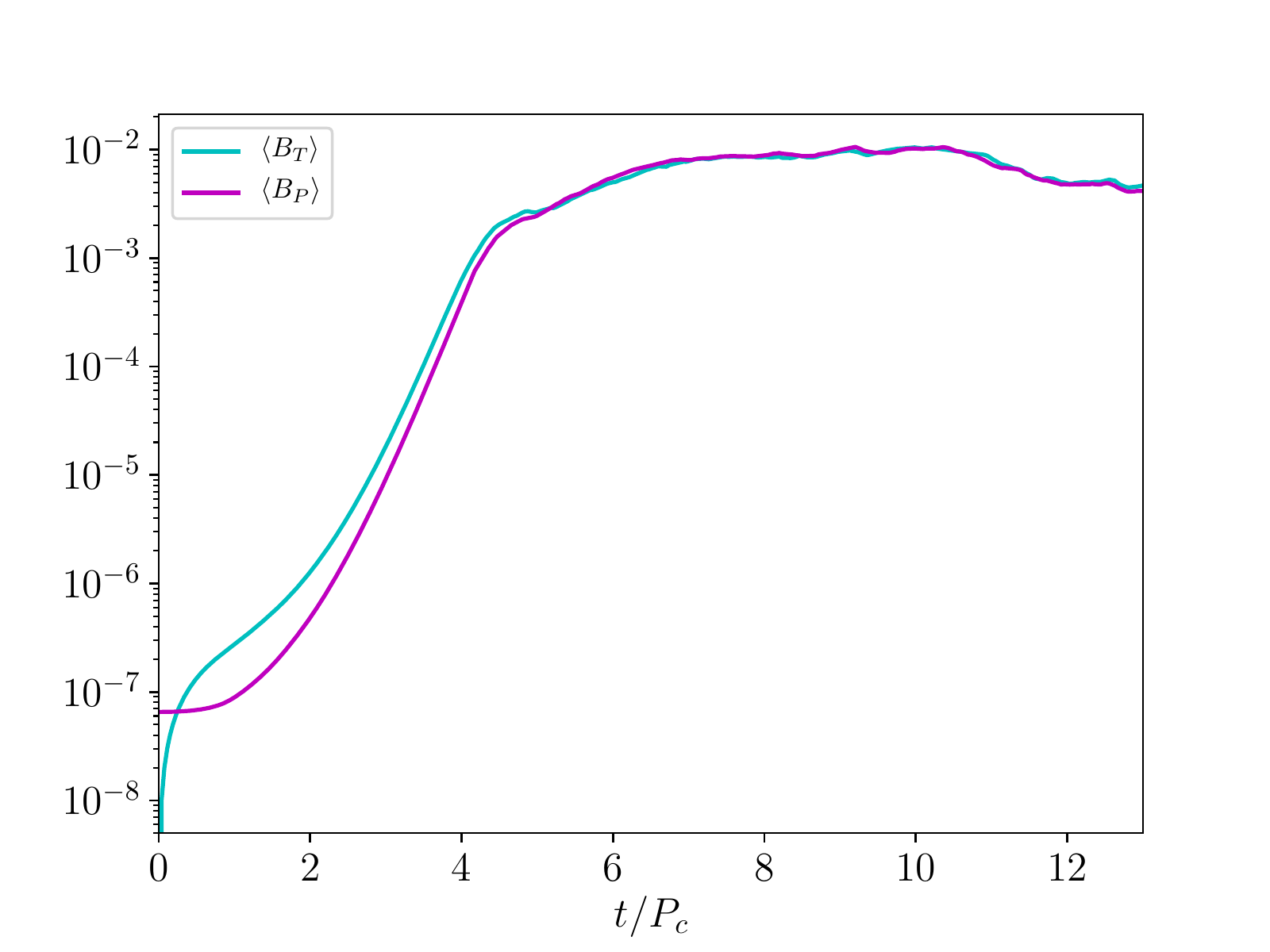}
\caption{The time evolution of the average values of the toroidal $B_T=\sqrt{B^\phi B_\phi}$ and poloidal $B_P=\sqrt{B^2 - B_T^2}$ components of the magnetic field.}
\label{fig:bt-bp}
\end{figure}

\begin{figure}
\centering
\includegraphics[width=.5\textwidth]{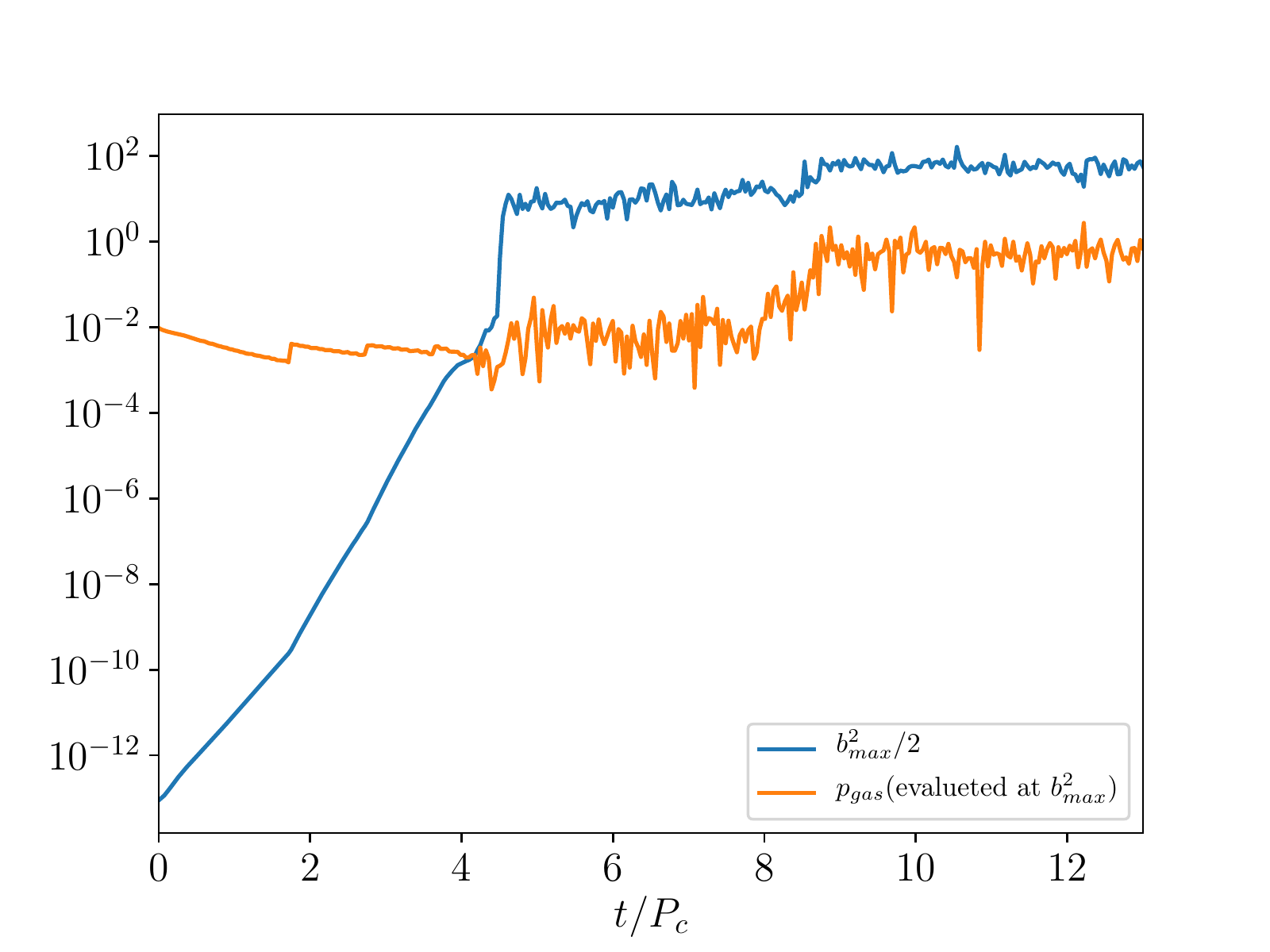}
\caption{Time evolution of $p_\mathrm{mag}=\frac{1}{2}b^2$ and $p_\mathrm{gas}\equiv p$, both evaluated where $p_\mathrm{mag}$ takes its maximum value.}
\label{fig:pmax}
\end{figure}

\begin{figure}
\centering
\includegraphics[width=.45\textwidth]{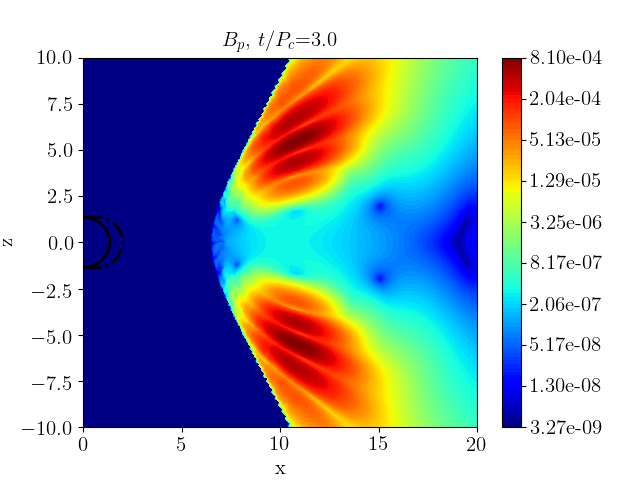}
\includegraphics[width=.45\textwidth]{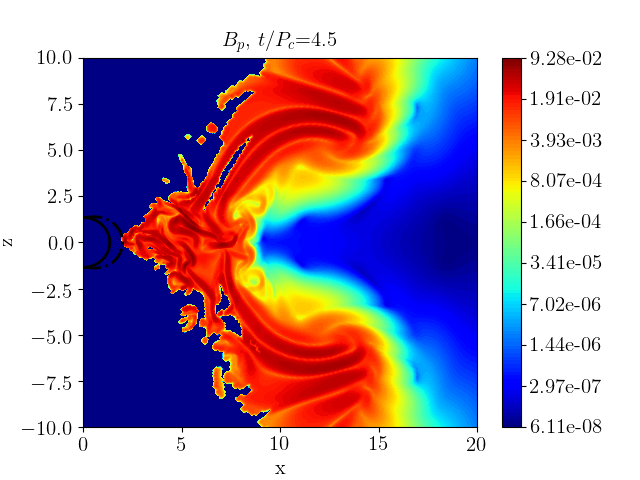}
\includegraphics[width=.45\textwidth]{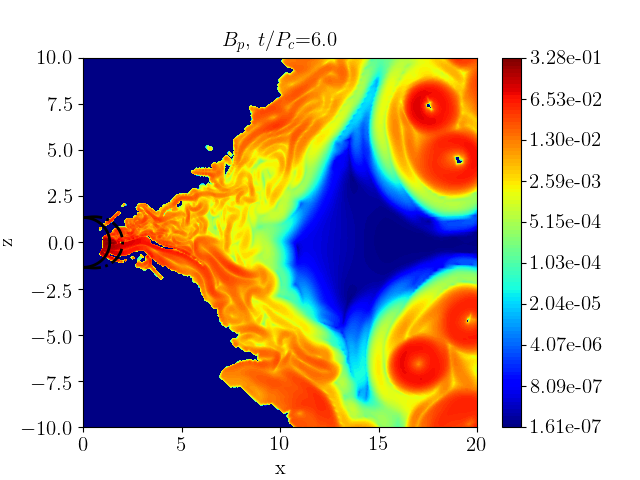}
\caption{Color maps of the poloidal magnetic field $B_P$ in logarithmic scale, for three different times $t/P_c$ Black lines near the origin are the contours of the black hole's ergosphere (dashed line) and horizon $r_h$ (solid line).}
\label{fig:Bp-map}
\end{figure}

In this section we show the results of $\alpha-\Omega$ dynamo simulations.
We start by defining the average on the disk of any quantity $f=f(r,\theta)$ as
\be
\langle f\rangle=\frac{\int_{r_1}^{r_2} \mathrm{d}r \int_{\theta_1}^{\theta_2}\mathrm{d}\theta \,\alpha \sqrt{\gamma}\, f}{ \int_{r_1}^{r_2} \mathrm{d}r\int_{\theta_1}^{\theta_2}\mathrm{d}\theta \,\alpha \sqrt{\gamma}},
\ee
where $r_1=4$, $r_2=30$, $\theta_1=\pi/3$, and $\theta_2=2\pi/3$. The assumption of limiting the average in this range is arbitrary, the reason behind this choice is to define a region of the disk in which the quantities of interest are significantly appreciable. Note that because of the presence of the lapse function $\alpha$ this is not a proper $3+1$ spatial averaging, though the above formula is the one most commonly adopted within the GRMHD community \citep{Porth:2019}. The time range of the simulations goes from 0 to $13P_c$, where $P_c=268$ is the initial central period.

Figure~\ref{fig:bt-bp} shows the time evolution of the average poloidal and toroidal components of the magnetic field in the Run1. We can see that a toroidal field immediately arises due to the $\Omega$ effect and, after a transient, the mean-field $\alpha-$dynamo starts as well and supports the exponential amplification of the two components up to $\sim4.5$~$ t/P_c$. This phase coincides with the kinematic regime studied by \citet{Bugli:2014}, as there is there is no noticeable feedback on the disk and the field grows following the normal modes of the dynamo, propagating towards the outer edge of the disk. During this linear phase the toroidal fields remain always stronger than the poloidal component. The new interesting aspect is represented by the situation around $t\simeq 4.5 P_c$, where the linear dynamo action saturates. As shown in the Figure~\ref{fig:pmax}, the transition occurs when there are regions where the gas pressure locally equals the magnetic one
\be
p_{mag}=\frac{1}{2}b^2=\frac{1}{2}(B^2-E^2),
\ee
and the sharp jump means that the most magnetized regions are no longer within the disk but start to form in the low-density atmosphere, where accretion is taking place. This corresponds to a change of slope in the growth of the magnetic field displayed in Figure~\ref{fig:bt-bp}: the dynamo action is less strong, though a secondary linear phase can still be recognized, and the values for the two magnetic field components are basically the same. After $t\simeq 8 P_c$ a second and definitive saturation stage has been reached, the dynamo amplification slowly begins to decrease, and the magnetic field approaches more or less to a constant value.

The three phases are more clearly apparent in Figure~\ref{fig:Bp-map}, where spatial maps of the (poloidal) magnetic field are presented (in logarithmic scale) at three different times. In the upper panel we are clearly still in the kinematic, linear phase of the dynamo. The magnetic field does not affect the disk shape, magnetic islands corresponding to the linear dynamo modes migrate towards the outer edge of the disk while growing in amplitude. In the middle panel the disk starts to be affected by the presence of the growing field and dynamo waves are dragged towards the black hole by the accretion. The accretion process is most probably triggered by MRI (see the discussion below), that also drives turbulent motions. In this dynamic regime, magnetic structures tend to form low-pressure vortices that drag matter away (bottom panel), that for high values ​​of the magnetic field can even evacuate the plasma locally (and safety floor density values can be required numerically, in order to limit this effect). The dynamo modes are barely visible during the phase of the secondary growth (third panel), and they seem to be localized only at the external boundary of the disk, where density and the dynamo term $\xi$ are lower (this point will be addressed in the next section).

\subsection{Dependence on the $\alpha-$dynamo number and on the quenching effect}

\begin{figure}
\centering
\includegraphics[width=.5\textwidth]{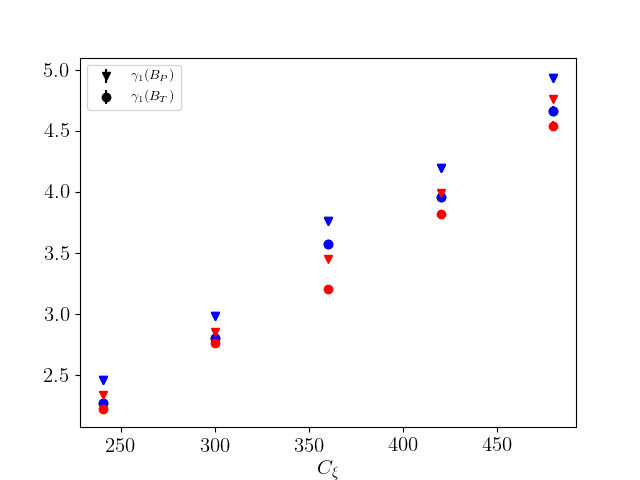}
\includegraphics[width=.5\textwidth]{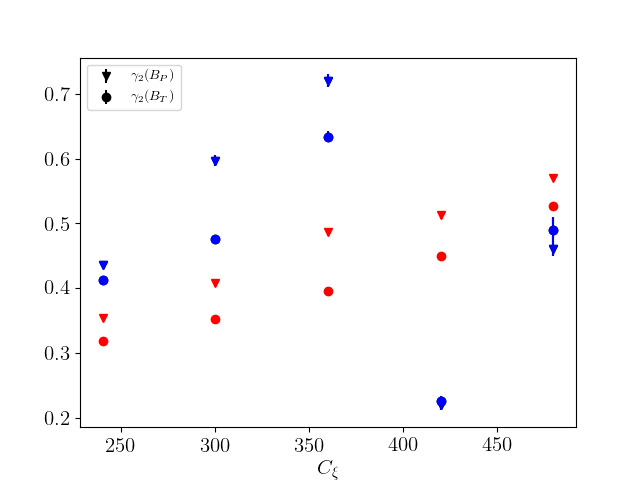}
\caption{Dependence of growth rates $\gamma_1$ and $\gamma_2$ on the dynamo number $C_\xi$. Triangles (circles) for the poloidal (toroidal) field components, blue (red) color for runs without (with) quenching.}
\label{fig:scatter}
\end{figure}

\begin{table*}
\centering
\small
\begin{tabular}{|c|c|c|c|c|c|c|c|c|c|}
\hline
& $\boldsymbol{\eta_\mathrm{max}}$& $\boldsymbol{\xi_\mathrm{max}}$ &$\boldsymbol{C_\xi}$ &$ \boldsymbol{C_\Omega}$ & $ \boldsymbol{\gamma_1(B_P)}$ &  $ \boldsymbol{\gamma_1(B_T)}$ &  $ \boldsymbol{\gamma_2(B_P)}$ &  $ \boldsymbol{\gamma_2(B_T)}$ \\
\hline
Run1 & $1.0\cdot10^{-3}$ & $3.0\cdot 10^{-2}$ & $3.6\cdot 10^2$ & $8.3\cdot 10^2$ &  $3.76\pm0.010$  & $3.57\pm0.01$  & $0.720\pm0.010$  & $0.634\pm0.008$ \\
Run2 & $1.0\cdot10^{-3}$ & $3.5\cdot 10^{-2}$ & $4.2\cdot 10^2$ & $8.3\cdot 10^2$ &  $4.190\pm0.030$  & $3.960\pm0.030$ & $0.220\pm0.009$  & $0.225\pm0.008$   \\  
Run3 & $1.0\cdot10^{-3}$ & $4.0\cdot 10^{-2}$ & $4.8\cdot 10^2$ & $8.3\cdot 10^2$ &  $4.930\pm0.030$  & $4.660\pm0.040$ & $0.460\pm0.010$  & $0.490\pm0.020$   \\  
Run4 & $1.0\cdot10^{-3}$ & $2.5\cdot 10^{-2}$ & $3.0\cdot 10^2$ & $8.3\cdot 10^2$ &  $2.980\pm0.010$ & $2.800\pm0.002$  & $0.597\pm0.008$  & $0.476\pm0.004$   \\ 
Run5 & $1.0\cdot10^{-3}$ & $2.0\cdot 10^{-2}$ & $2.4\cdot 10^2$ & $8.3\cdot 10^2$ &  $2.460\pm0.010$  & $2.267\pm0.009$ & $0.436\pm0.002$  & $0.412\pm0.005$   \\ 
Run1q & $1.0\cdot10^{-3}$ & $3.0\cdot 10^{-2}$ & $3.6\cdot 10^2$ & $8.3\cdot 10^2$ &  $3.450\pm0.010$  & $3.200\pm0.020$ & $0.486\pm0.005$  & $0.396\pm0.003$ \\
Run2q & $1.0\cdot10^{-3}$ & $3.5\cdot 10^{-2}$ & $4.2\cdot 10^2$ & $8.3\cdot 10^2$ &  $3.990\pm0.020$  & $3.820\pm0.030$ & $0.513\pm0.006$  & $0.449\pm0.004$   \\  
Run3q & $1.0\cdot10^{-3}$ & $4.0\cdot 10^{-2}$ & $4.8\cdot 10^2$ & $8.3\cdot 10^2$ &  $4.760\pm0.020$  & $4.540\pm0.040$ & $0.570\pm0.003$  & $0.526\pm0.003$   \\  
Run4q & $1.0\cdot10^{-3}$ & $2.5\cdot 10^{-2}$ & $3.0\cdot 10^2$ & $8.3\cdot 10^2$ &  $2.850\pm0.010$  & $2.760\pm0.010$ & $0.408\pm0.006$  & $0.352\pm0.004$   \\ 
Run5q & $1.0\cdot10^{-3}$ & $2.0\cdot 10^{-2}$ & $2.4\cdot 10^2$ & $8.3\cdot 10^2$ &  $2.336\pm0.007$  & $2.220\pm0.010$ & $0.354\pm0.005$  & $0.318\pm0.003$   \\ 
\hline
\end{tabular}
\caption{Initialization parameters for the various runs and growth rates in the two phases. The reported dynamo numbers refer to their maximum value. The initial plasma beta is $\beta = 10^9$ for all runs.}
\label{tab:run-models}
\end{table*}

We now investigate the dependence of the results on the $\alpha-$dynamo number $C_\xi$ and on the employment of an explicit quenching effect (see below). The list of runs with the corresponding parameters is reported in Table~\ref{tab:run-models}. In particular, Run2 and Run3 are characterized by increasing values of $\xi$ and $C_\xi$ with respect to our reference Run1 values, whereas Run4 and Run5 by decreasing values of the same parameters. Figure~\ref{fig:scatter} shows the dependence of the exponential growth rates of the kinematic ($\gamma_1$) and dynamic ($\gamma_2$) phases, respectively, with the dynamo number $C_\xi$. Growth rates are measured for both the poloidal field component (blue triangles) and for the toroidal one (blue circles), values also reported in Table~\ref{tab:run-models}. The red symbols indicate the corresponding quantities for simulations where quenching is active (labeled with a 'q' in the table of runs).

We observe that the rates $\gamma_1$ corresponding to the kinematic phase follow a linear trend, as expected, whereas the rates $\gamma_2$, corresponding to the phase where accretion affects the dynamo modes, show an unexpected drop at high values ​​of $\xi$ (blue symbols). This may be due to the rapid growth of the magnetic field, leading to values able to modify the fluid equilibrium itself. This seems to prevent, or at least to lower, a subsequent amplification, as if a saturated state has been reached.

In order to limit the dynamo action and to obtain a more regular growth it is possible to adopt a technique, used in purely kinematic dynamo models to simulate dynamic effects, that is to impose an explicit quenching in the dynamo term in situations when the field becomes comparable to the equipartition value \citep[e.g.][]{Brandenburg:2005}. This is obtained by introducing, \emph{locally} at any point, the replacement
\be
\xi\rightarrow\frac{\xi}{1+B^2/B^2_\mathrm{eq}},
\ee
where we have considered an equipartition turbulent field defined as a given fraction of the thermal pressure \citep{Shakura:1973}, $B_\mathrm{eq}^2=\bar{\alpha}_\mathrm{disk}p$, with $\bar{\alpha}=0.1$. This value is the one most commonly used to model the turbulent magnetic stresses in disks \citep{King:2007}. 

\begin{figure}
\centering
\includegraphics[width=.5\textwidth]{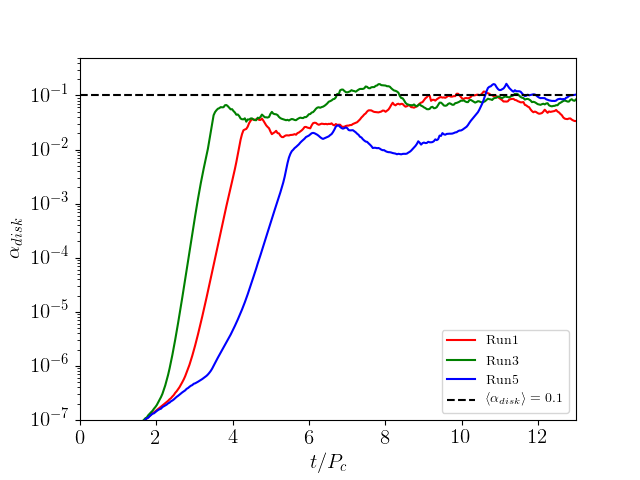}
\caption{The quantity ${\alpha}_\mathrm{disk}$ defined in Eq.~(\ref{eq:alpha_disk}) as a function of time, for three runs with different dynamo number $C_\xi$.}
\label{fig:alpha}
\end{figure}

Here we want to check whether our GRMHD simulations without quenching lead to a turbulent state with fluctuations of the required intensity. We thus compare this value with the coefficient ${\alpha}_\mathrm{disk}$ defined by the averages
\be
{\alpha}_\mathrm{disk}=\frac{\langle W \rangle}{\langle p+p_\mathrm{mag} \rangle},
\label{eq:alpha_disk}
\ee
where, for rotating disks in GRMHD \citep[e.g.][]{Bugli:2018}  
\be
W=[(w+p+b^2)\delta u^r\delta u^\phi-b^rb^\phi]\sqrt{\gamma_{\phi\phi}}\sqrt{\gamma_{rr}}
\ee
is the $r, \phi$ component of the stress tensor of fluctuations, based on the variations of the relevant 4-velocity components (compared to the equilibrium state at $t=0$) and on the growing fluctuations of the magnetic 4-vector components (negligible at $t=0$). As shown in Figure~\ref{fig:alpha}, the choice of the value $\bar{\alpha}=0.1$ for the quenching is reasonable, since all runs saturate towards average values of ${\alpha}_\mathrm{disk}$ with this value, or slightly lower. This means that the introduction of the quenching effect is not expected to affect the overall dynamics, but just to limit the growth of the field in localized, critical zones.


The five runs have been repeated with identical parameters and the addition of the quenching effect, labelled as Run1q-Run5q in Table~\ref{tab:run-models}. As expected, the dynamo growth rates are basically unchanged in the quasi-kinematic phase. However, now the rates $\gamma_2$ appear to grow linearly with $C_\xi$, exactly as rates $\gamma_1$, even in the phase where turbulence and accretion are present (see the red symbols in Figure~\ref{fig:scatter}, for both $\gamma_1$ and $\gamma_2$). Notice that below $C_\xi = 400$ the $\gamma_2$ values are lower than the corresponding cases without quenching, as expected, but higher above that value, where however the blue data looked pathological since no regular trend was followed. Four additional runs with $C_\xi$ increasing from $\simeq 750$ up to $\simeq 1800$ have also been performed, again in presence of the quenching term. The linear trend for $\gamma_2$ is less evident than what shown in Figure~\ref{fig:scatter}, though we find a final value $\gamma_2\simeq 1.5$, that is more or less what one would expect from a linear extrapolation.

\begin{figure}
\centering
\includegraphics[width=.5\textwidth]{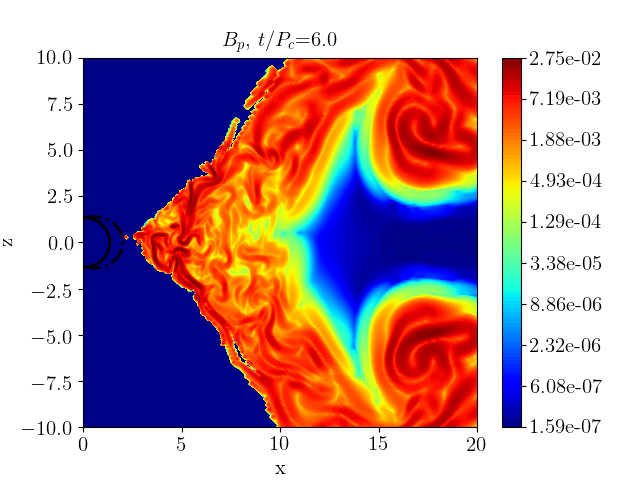}
\caption{Map of the poloidal field (in logarithmic scale) for $t=6P_c$ (Run1q, with quenching), to be compared with the third panel of Figure~\ref{fig:Bp-map} (Run1, without quenching).}
\label{fig:Bp-t=6-q}
\end{figure}

These results show that the dynamo appears to be the main mechanism for amplifying magnetic fields even during the phase in which the disk starts to lose mass, which is accreting onto the black hole. Furthermore, when quenching is activated, since a lower magnetic field is present, the formation of vortices evacuating the plasma is inhibited and the dynamo structures evolve more smoothly even in a turbulent environment, as clearly shown in Figure~\ref{fig:Bp-t=6-q}.

\begin{figure}
\centering
\includegraphics[width=.5\textwidth]{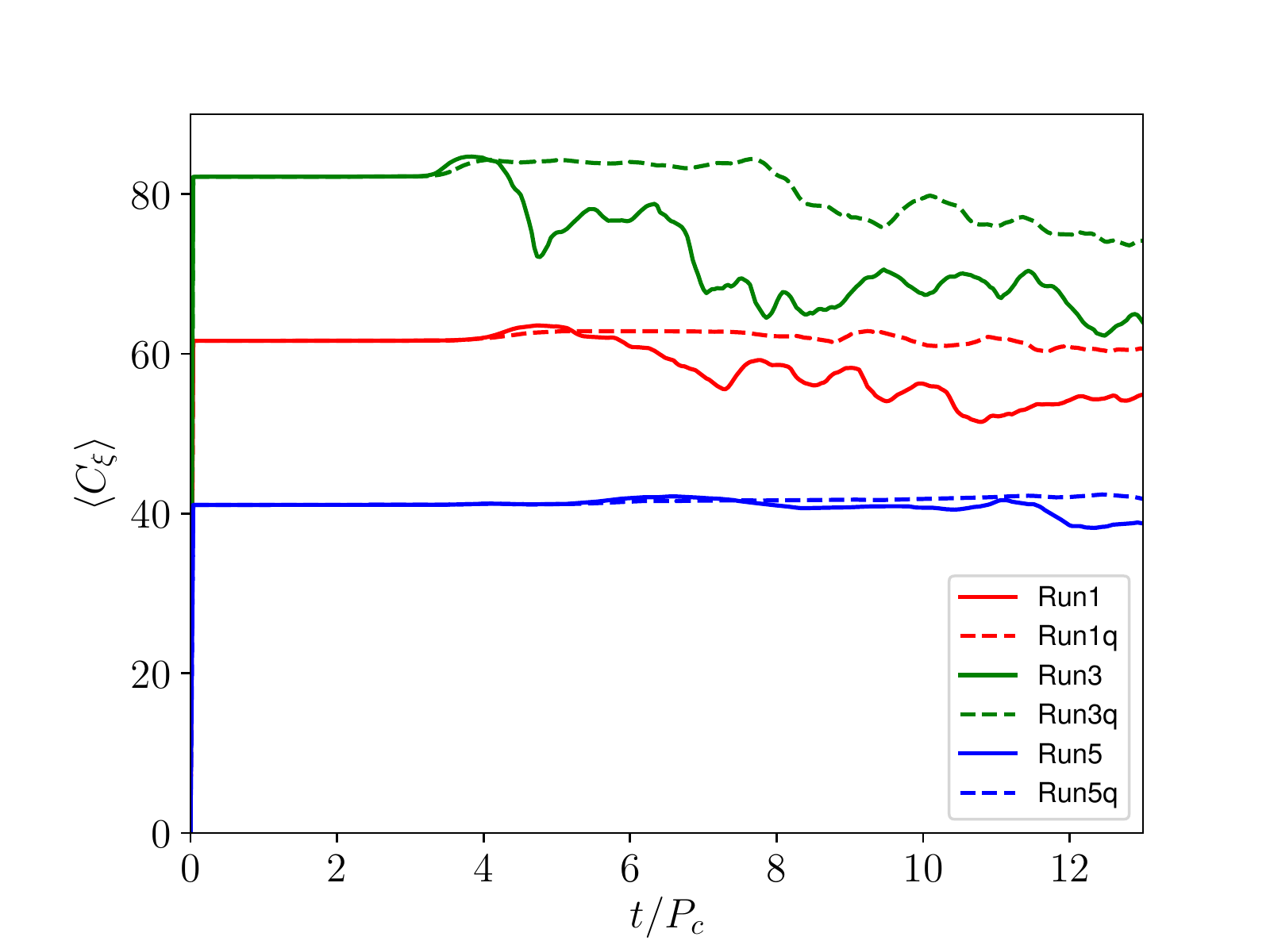}
\caption{Time series of the dynamo number $C_\xi$ averaged over the whole disk.}
\label{fig:C_xi}
\end{figure}

In order to better clarify why the secondary growth rate $\gamma_2$ is lower than the corresponding $\gamma_1$, we plot in Figure~\ref{fig:C_xi} the time series of the spatial averages of the $C_\xi$ dynamo number. We clearly see that, when the accretion begins and the first saturation phase starts, the quantity decreases for all runs without the quenching term and for Run3q as well. This is due to the fact that during accretion the density can be modified substantially, and $< C_\xi >$ as well, and this effect is stronger for higher magnetization levels. When the quenching term is present both magnetization and turbulence are generally lower, the presence of the low-density vortices is avoided, and the overall dynamics is certainly more regular. In any case, especially for runs with quenching, the lower values of $\gamma_2$ cannot be attributed to correspondingly smaller values of $<C_\xi>$, while migration of the plasma near the border of the disk, where $C_\xi$ is reduced, appears to be more important. In addition, when the density is low, the other dynamo term, $C_\Omega$, gets larger values, and for a constant $C_\xi$ the overall growth rates are expected to be reduced \citep[see Table~1 in][]{Bugli:2014}.

\begin{figure}
\centering
\includegraphics[width=.5\textwidth]{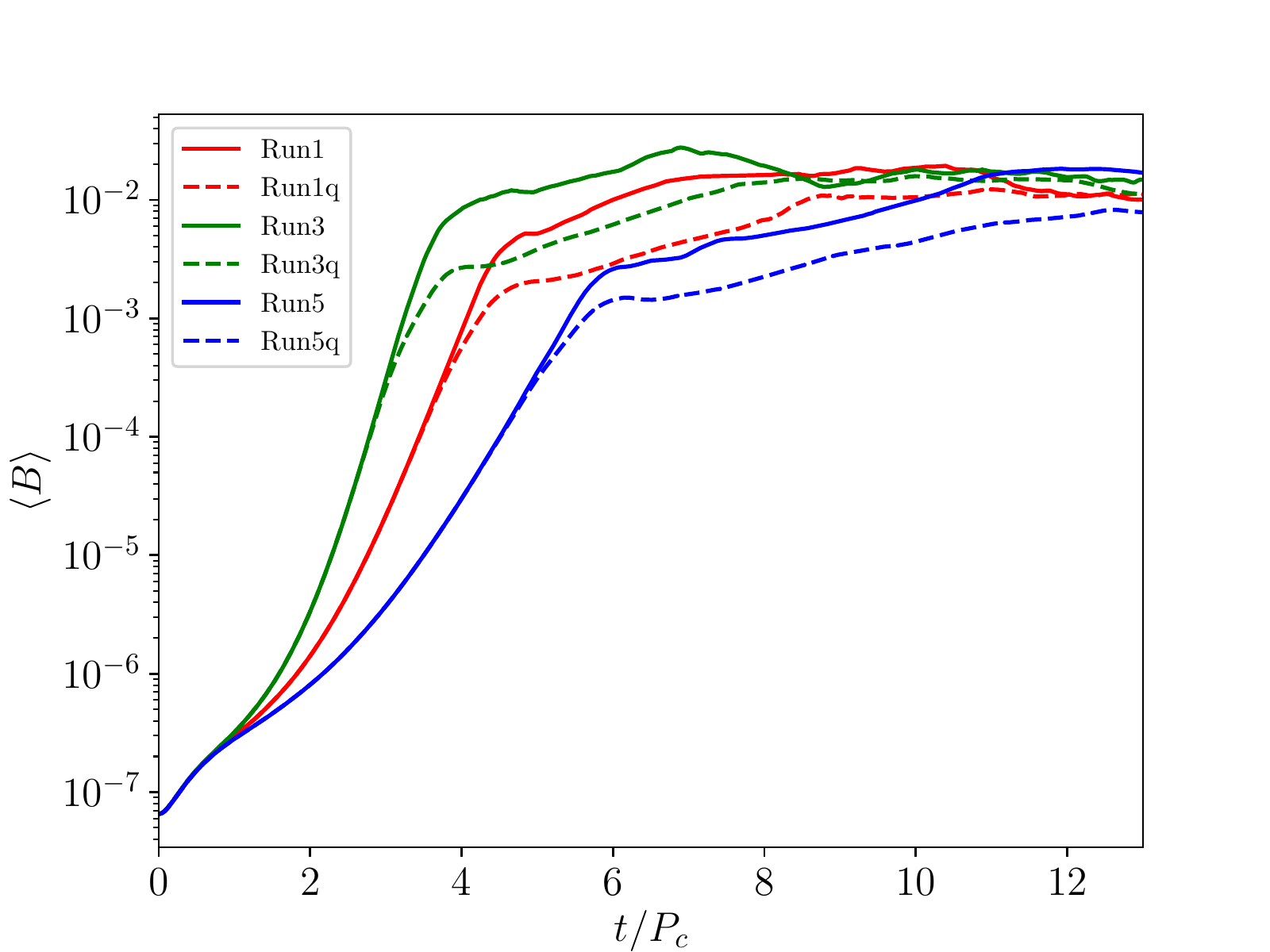}
\caption{Time evolution of the averaged intensity of the magnetic field, for three runs without and with quenching.}
\label{fig:b_av}
\end{figure}

\begin{figure}
\centering
\includegraphics[width=.5\textwidth]{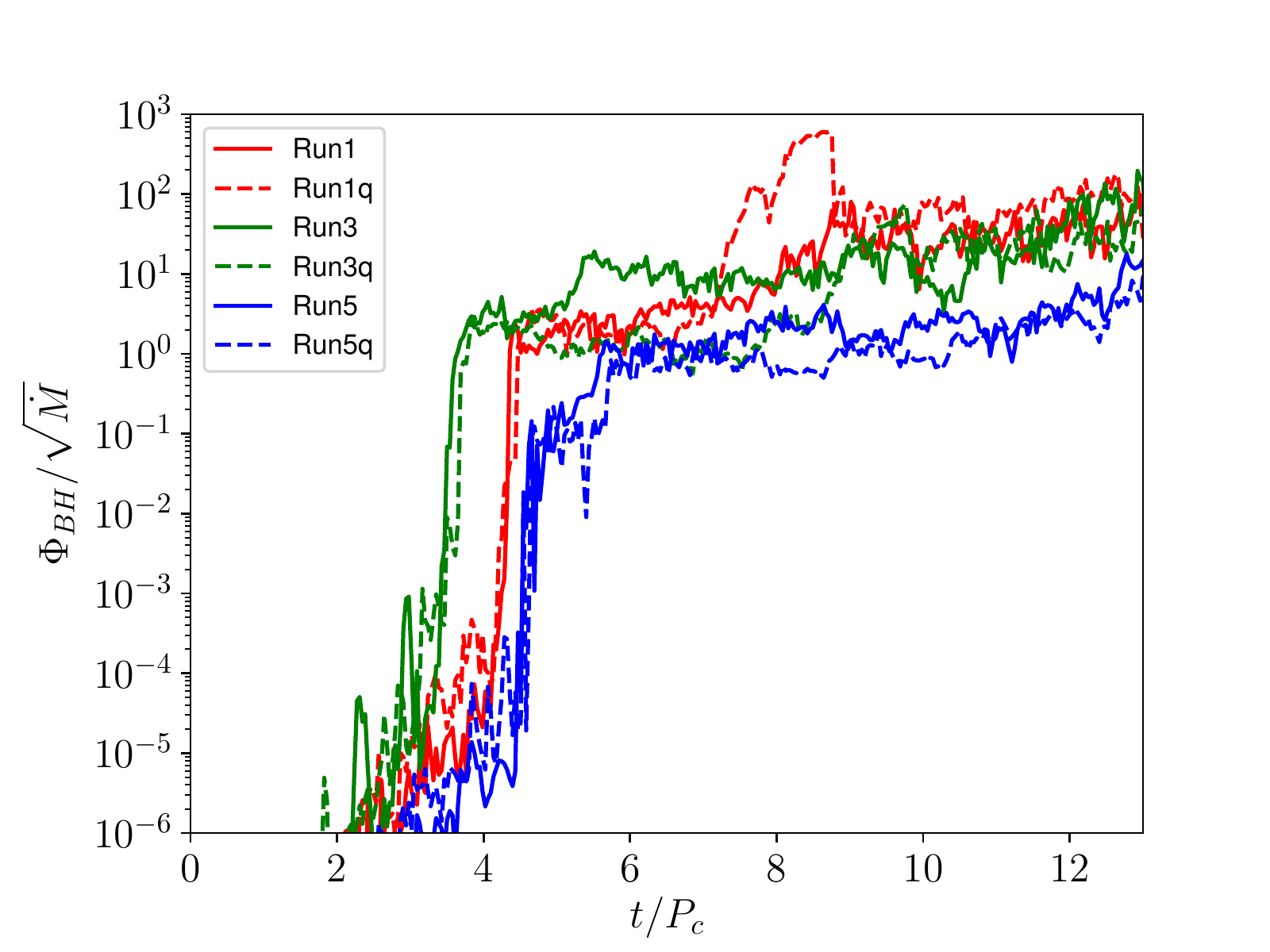}
\caption{Time evolution of the magnetic flux $\Phi_\mathrm{BH}$ penetrating the horizon, divided by the accretion rate, for the same six runs as in Figure~\ref{fig:b_av}.}
\label{fig:bflux}
\end{figure}

A very interesting result is that the value of $C_\xi$ and the presence of quenching do not affect too much the value of the quantities in the final saturation stage, as can be seen in Figure~\ref{fig:b_av}, where the growth of the average strength ​​of the magnetic field is plotted for our six reference runs. The initial kinematic growth and also the \emph{first} saturation phase clearly depend on $C_\xi$, as the fastest growing mode occurs at a wavenumber $\sim \xi/\eta$ \citep{Brandenburg:2005}, with smaller scales triggered by MRI leading to a turbulent cascade towards the dissipative scales. However, the \emph{second} and final saturation phase looks approximately independent on the $C_\xi$ value and on the presence of quenching (all curves tend to the same value of $<B>$ within a factor of $\simeq 3$), hence we deem that the accretion dynamics and the reached equipartition with the fluid component play a major role at this final stage (see the similar behavior of ${\alpha}_\mathrm{disk}$).

For completeness, the magnetic flux threading one hemisphere of the black hole horizon, $\Phi_\mathrm{BH}$, has been evaluated, a quantity which is very important because the rotational energy extracted, the Blandford-Znajek power, is proportional to its square  \citep{Blandford:1977,Tchekhovskoy:2011}. This is defined as
\be
\Phi_\mathrm{BH} = \frac{1}{2}\,2\pi \int_0^{\pi}|{B^r}| \sqrt{\gamma} \, \mathrm{d}\theta ,
\ee
to be evaluated at the outer event horizon $r_h$. Note that throughout the literature there are two different definitions of the magnetic field components in terms of the dual of the Faraday tensor: one as $B^i = F^{\star i0}$ \citep[e.g.][]{McKinney:2004}, the other as $B^i = - n_\mu F^{\star\mu i} = \alpha F^{\star 0 i} $ \citep[e.g.][]{DelZanna:2007}, where $n^\mu$ is the Eulerian observer unit vector (only this second one is a proper spatial projection according to the $3+1$ splitting). Figure~\ref{fig:bflux} describes, for our six reference runs, the time evolution of the so-called \emph{MAD parameter} $\phi = \Phi_\mathrm{BH} /\sqrt{\dot{M}}$, that is the above quantity normalized to the square root of the mass accretion rate. This quantity is commonly used to discriminate SANE evolution models from MAD ones \citep[e.g.][]{Porth:2019}, depending whether its maximum value is below or above the threshold of $\sim 15$. In our runs, in order to reach the typical SANE values we have of course to wait for the saturation of the first magnetic field growth, for which we find $\phi\sim 1$. At later times we observe a persistent slow growth, due to the secondary dynamo action, and for Run1 and Run3 even the MAD phase seems to be reached, with final values in the range $\phi\sim 50 - 100$.

\subsection{On the magnetorotational instability}

In the non-linear regime it is interesting to investigate whether MRI is capable of affecting the dynamo action or, more generally, of changing substantially the structure of the magnetic field. 

It is custom to define the so-called MRI quality factor, a parameter that allows to establish if a given simulation is able to resolve the characteristic MRI wavelength $\lambda_\mathrm{MRI}$, precisely by measuring the number of cells contained in $\lambda_{\mathrm{MRI}}$, for a given direction. In our axisymmetric case the important quality factor is the one along the direction $\theta$, hence here we define
\be
Q_\theta = 
\frac{\lambda_{\mathrm{MRI},\theta}}{ \sqrt{\gamma_{\theta\theta} } \Delta \theta }=
\frac{ 2\pi |v_A^\theta | }{\Omega \, \sqrt{\gamma_{\theta\theta} } \Delta \theta },
\ee 
where $v_A^\theta = B^{\theta}/\sqrt{w+B^2}$ is the $\theta$-component of the relativistic Alfvén velocity (here neglecting the contribution by the electric fields). Values of $Q_\theta>6-8$ have been shown to be necessary to capture locally the linear growth of MRI, while a threshold of $Q_\theta>10$ can capture its nonlinear growth \citep{Noble:2010,McKinney:2012,Hogg:2018}. 

\begin{figure}
\centering
\includegraphics[width=.5\textwidth]{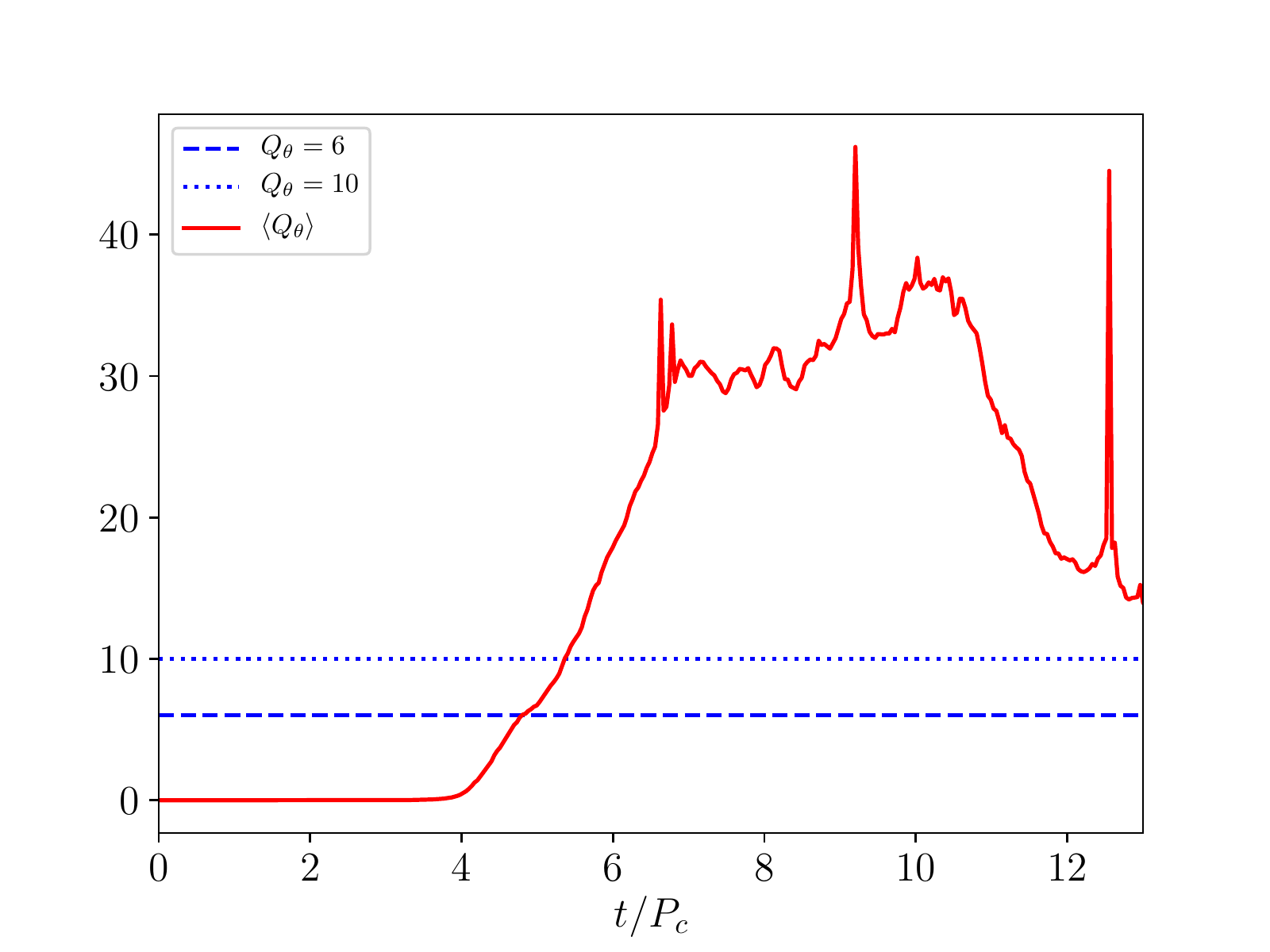}
\caption{Time dependence of the averaged MRI quality factor $Q_\theta$, for Run1 parameters. The two horizontal lines are the thresholds for resolving the linear (dashed line) and nonlinear (dotted line) phases.}
\label{fig:qtheta}
\end{figure}

Figure~\ref{fig:qtheta} shows the time evolution of the factor $Q_\theta$, averaged over the whole disk, in the case of Run1. Apparently the MRI instability could be resolved during the dynamical phase and therefore, if present, it would be expected to play a role in the growth of the magnetic field, competing with the ongoing dynamo process.

\begin{figure}
\centering
\includegraphics[width=.5\textwidth]{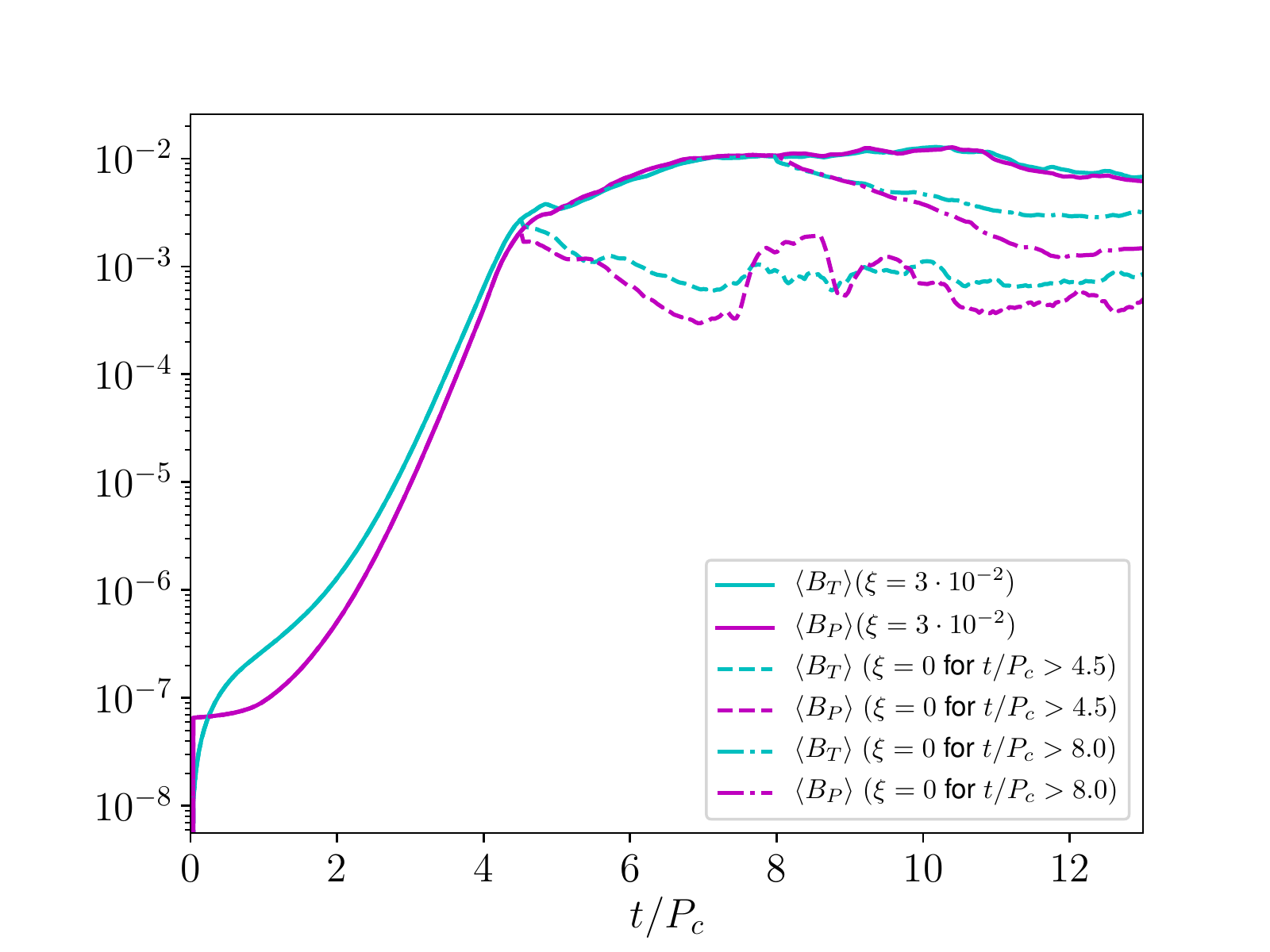}
\caption{Growth of the magnetic field components for Run1 and for two additional runs with the same parameters but switching off the dynamo ($\xi=0$) for $t/P_c > 4.5$ (dashed lines) and for $t/P_c > 8.0$ (dot-dashed lines).}
\label{fig:stop_dyn}
\end{figure}

To investigate this aspect we have re-executed Run1 twice, the first time by setting $\xi=0$ for $t/P_c > 4.5$, at the end of the first linear stage, and the second one for $t/P_c > 8.0$, just before the final saturation after the second linear phase. As we can see in Figure~\ref{fig:stop_dyn}, the magnetic field components immediately start to decrease after the dynamo has been switched off, in both cases. Notice that the poloidal component is the one suffering the fastest decrease, the toroidal one is probably still supported by a residual amplification due to rotation (a purely $\Omega$ effect).

This result means that the dynamo action is the main driver for magnetic field enhancement in all phases, whereas MRI, which is surely present and resolved numerically in our simulations, seems to be responsible mainly for triggering turbulence and driving the accretion. Amplification of magnetic fields by MRI is instead less strong than the one due to the mean-field dynamo (we recall that our simulations are 2D axisymmetric), and its effect is only visible in the final saturation stage, where a faster decrease due to dissipation is inhibited. We conclude by saying that the adopted resistivity value of $\eta=10^{-3}$ is still above the level of numerical dissipation, which is estimated to be $\eta  \sim 10^{-4}$ for the resolution employed.

\subsection{Comparison with Sgr A* radio emission}

Our GRMHD models based on the dynamo action will be used here to infer the synthetic emission by the magnetized plasma of the accreting matter and to compare it with observational data. The most straightforward targets are obviously the two sources observed by EHT, Sgr A* and M87*, i.e. the cores containing the super-massive black holes of our Galactic center and of the elliptical galaxy M87. In the latter case the very first image of the emission from the regions around a black hole's event horizon has been recently taken \citep{EHTCollaboration:2019a}, whereas at the moment work is in progress to reduce data in the case of Sgr A*.

Since our simulations are more focussed on the plasma dynamics occurring inside the disk, with the magnetic field growing from initial seed values, we choose here to compare our model with Sgr A* data, as in the case of M87* a substantial fraction of the emission is known to come from the polar jet. The structure of Sgr A* is rather uncertain because the source is hidden by optically thick interstellar medium that surrounds it. For this reason models with or without a jet have been built over the years to infer its emission properties \citep{Moscibrodzka:2009,Moscibrodzka:2013,Moscibrodzka:2014}. 

The radio emission in the mm band of the Sgr A* spectrum can be modelled by the radiation produced by  thermal synchrotron-emitting relativistic electrons in a ADAF/RIAF (\emph{Advection-Dominated Accretion Flow and Radiatively Inefficient Accretion Flow}) model. According to the theory, most of the energy is stored in the thick pressure-supported disk and advected inwards with high speed and efficiency. The large scale-height and accretion velocity makes the density low, the gas cooling time long compared to advection times (the temperature of proton remains high, $T_p \sim 10^{11} - 10^{12}$~K), and the plasma is optically thin \citep{Narayan:1995a,Narayan:2008}.

The total emissivity per unit frequency $j_\nu$ is given by \citep{Leung:2011}
\be
\label{eq:magneto_emiss}
j_\nu=n_e\frac{\sqrt{2}\pi e^2 \nu_s}{6\Theta_e^2 c}X\exp\bigl(-X^{1/3}\bigr),
\ee
where $X=\nu/\nu_s$ (with $\nu\gg\nu_s$ in the above approximation), $\nu_s=(2/9)\nu_c\Theta_e^2\sin{\theta}$ is a characteristic frequency threshold ($\theta$ is the \emph{pitch angle} of the particle), and $\nu_c=eB/2\pi m_e c$ is the cyclotron frequency. Moreover $\Theta_e=kT_e/m_ec^2$ is the normalized electron temperature, and $m_e$ and $n_e$ are respectively the electron mass and numerical density. For an optically thin source, like Sgr A*, the spectral luminosity is simply defined by
\be
\label{eq:flux}
L_\nu = 2\pi \int_{r_h}^{r_2} \mathrm{d}r \int_{\theta_1}^{\theta_2} \mathrm{d}\theta \sqrt{\gamma}\, j_\nu,
\ee
and the observed flux would be $F_\nu = L_\nu/4\pi d^2$, where $d=7.86$~kpc is the estimated distance of the Galactic Center \citep{Boehle:2016}.

In RIAF systems, electrons are cooled by synchrotron, inverse-Compton, and bremsstrahlung emission losses, but ions maintain their high temperatures due to inefficient thermalization, thus a two-temperature plasma where the electron temperature is much lower than that of protons, $T_e\ll T_p$, is usually assumed \citep{Narayan:1995}. The ratio between proton and electron temperatures is assumed to be given by the expression \citep{EHTCollaboration:2019}
\be
\frac{T_p}{T_e} \equiv R  =R_\mathrm{high}\frac{\beta^2}{1+\beta^2}+\frac{1}{1+\beta^2},
\label{eq:Tratio}
\ee
where $\beta=p/p_\mathrm{mag}$ and $R_\mathrm{high}$ is a parameter that takes into account the electron-to-proton coupling in the regions of the disk where $\beta$ is high, $R\to R_\mathrm{high}$ for $\beta\gg 1$ \citep{Moscibrodzka:2016}.

In order compute the above emission quantities it is necessary to appropriately convert all quantities from code units to the CGS system, hence
\be
n_e=n_p= (\rho/\rho_0) n_0,
\ee
where we recall that $n_0$ is a free parameter providing the number density at the disk center, the magnetic field in expressed in units $B_0 = \sqrt{4\pi\rho_0 c^2}$, the normalized proton temperature is defined as
\be
\Theta_p = \frac{k T_p}{m_p c^2} = \frac{p/p_0}{\rho / \rho_0},
\ee
whereas $\Theta_e$ can be inferred from Eq.~(\ref{eq:Tratio}). Reference units for length and time are determined once the mass of the black hole has been assigned, for Sgr A* we assume $M_\mathrm{BH} =4.02 \times 10^6 M_\odot$ \citep{Boehle:2016}.

\begin{figure}
\centering
\includegraphics[width=.49\textwidth]{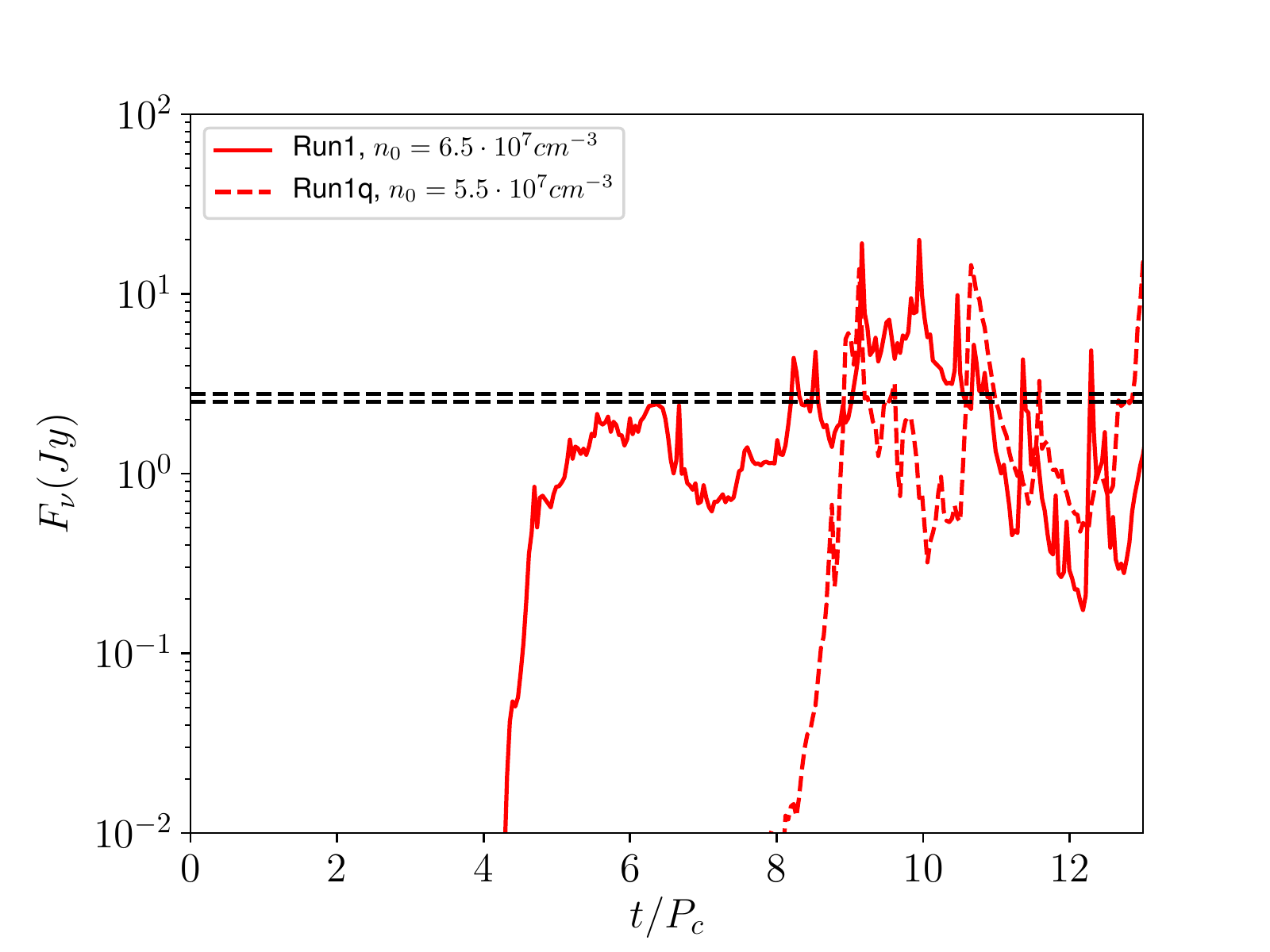}
\caption{Time evolution of the synthetic flux $F_\nu$ computed on top of Run1 (solid line), and of Run1q (dashed line). The horizontal dashed lines represent the observed value at for the reference frequency $\nu = 230$~GHz.}
\label{fig:flux-jansky}
\end{figure}

\begin{figure}
\centering
\includegraphics[width=.49\textwidth]{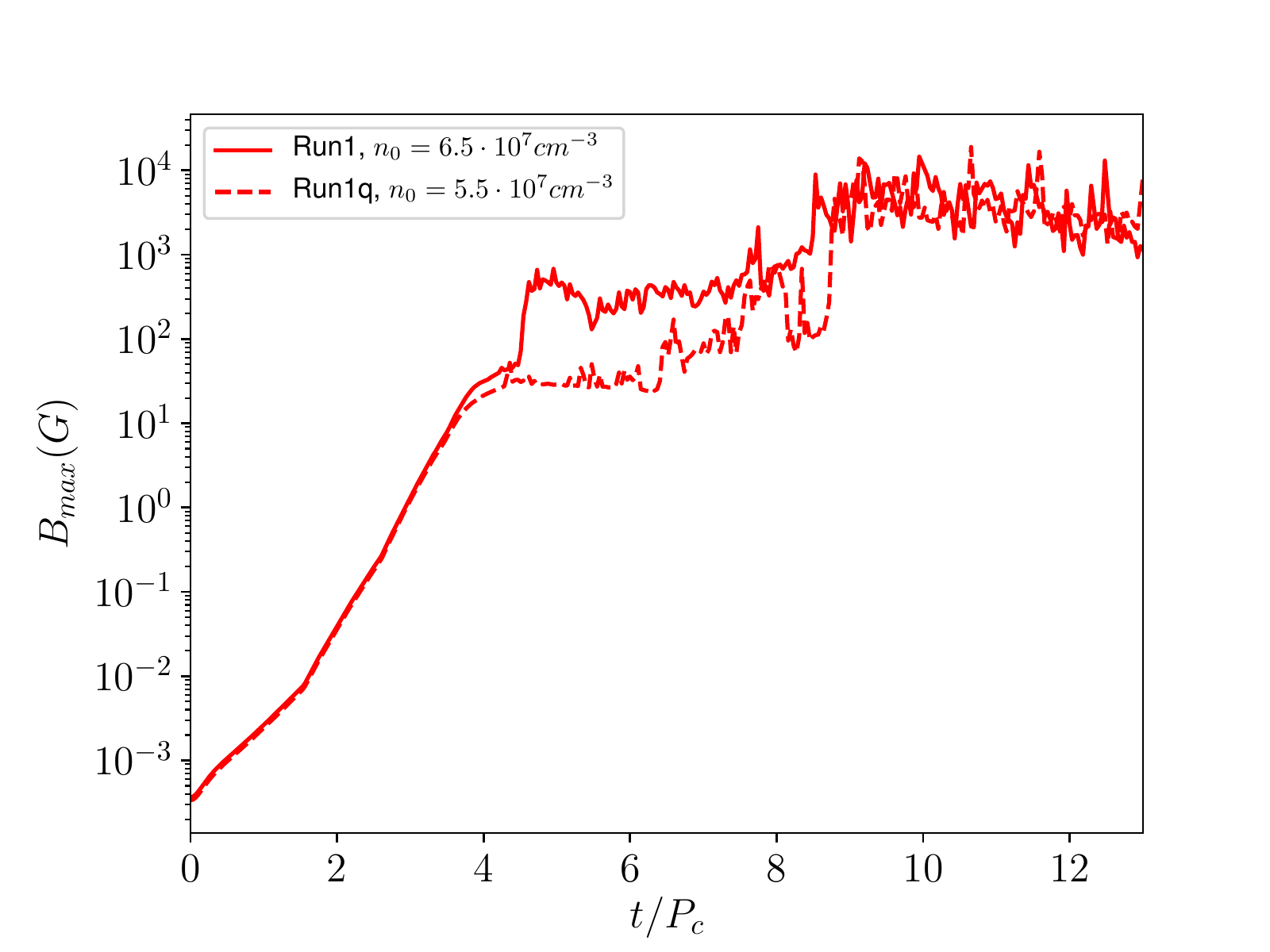}
\includegraphics[width=.49\textwidth]{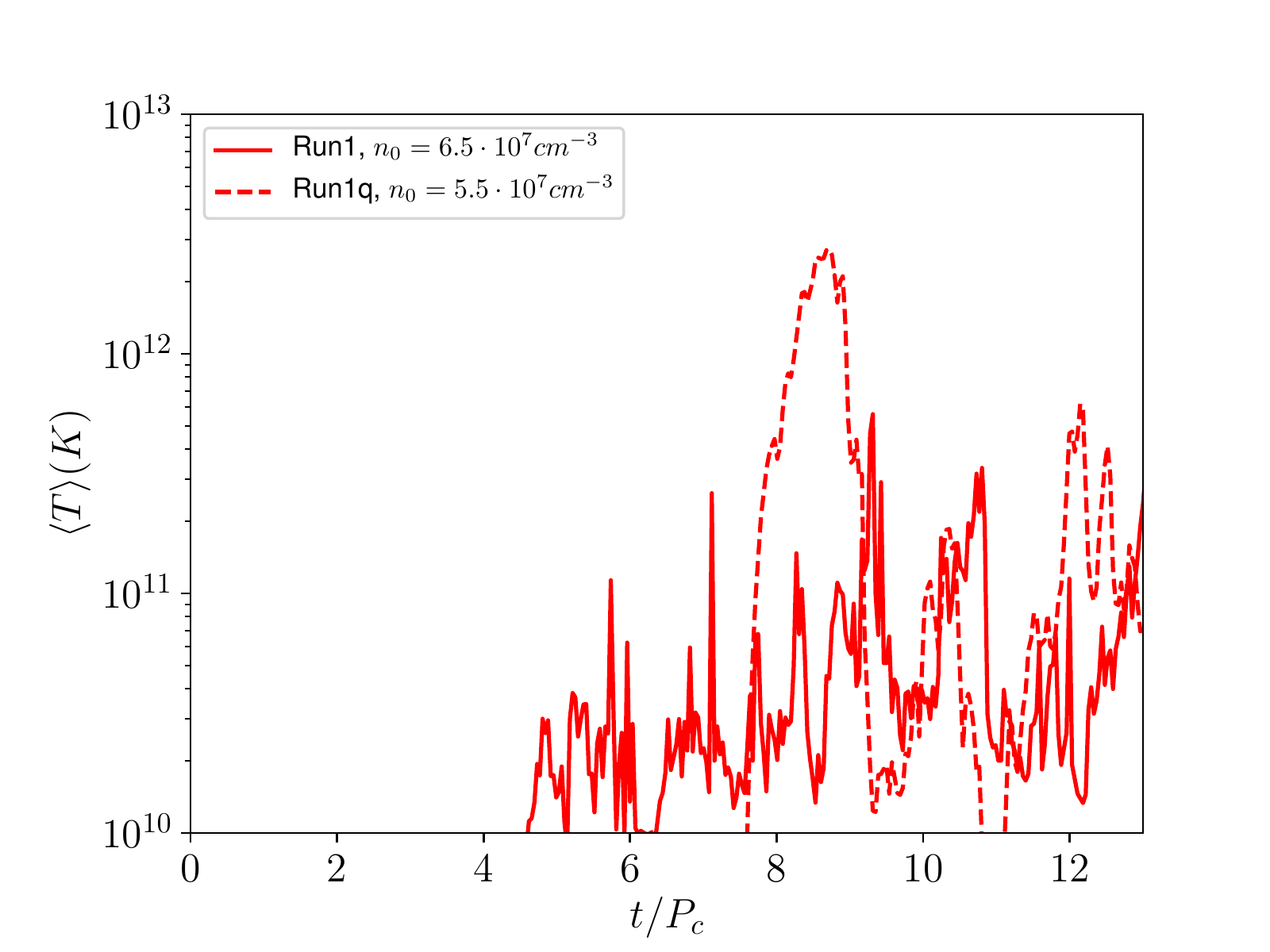}
\caption{Time evolution of the maximum value of the magnetic field strength (upper panel) and of the average electron temperature (lower panel) for Run1 (solid line), and of Run1q (dashed line).}
\label{fig:B}
\end{figure}

%

In the following, we show that our model (we choose Run1 and Run1q parameters) can predict that the initial magnetic field ($\sim10^{-4} ~\text{G}$) is amplified by the mean-field dynamo up to a level able to explain the observed flux in the millimetric wavelengths. The three free parameters left, namely $\nu$, $R_{high}$ and $n_0$, are chosen as follows: $\nu$ is set to 230~GHz, the same used for M87*, $R_{high}$ is fixed to 20, a reasonable value in the disk and $n_0$ is chosen as the value that best reproduces the observed flux of $2.64\pm 0.14$~Jy at 230~GHz \citep{Moscibrodzka:2013,Moscibrodzka:2014,EHTCollaboration:2019a}. Let us look at Figure~\ref{fig:flux-jansky}. The predicted synthetic flux follows the average magnetic field trend, increasing until it reaches a quasi-stationary value. We perform a time average in the range $t/P_c = [7 - 13]$, obtaining a flux of $2.59$ Jy consistent with the observations when assuming $n_0 = 6.5\cdot 10^{7} $ cm$^{-3}$ for Run1 and $n_0 = 5.5\cdot 10^{7} $ cm$^{-3}$ for Run1q. 

In Figure~\ref{fig:B} (upper panel) we show the time evolution of the maximum value of the magnetic field in the disk, reaching $B_\mathrm{max}\simeq 3-4$~kG at the saturation of the dynamo process. At the same time, the average electron temperature reaches $<T_e>$ just below $10^{11}$~K (lower panel). Overall these quantities result in agreement with the values ​​obtained in the simulations of \citet{Moscibrodzka:2009,Moscibrodzka:2013}.


\section{Summary and conclusions}

In the present paper we have investigated, for the first time by means of \emph{non-ideal} axisymmetric GRMHD simulations, the mean-field dynamo process operating in thick accretion disks around black holes, in the fully nonlinear regime. This work can be seen as a follow up of our previous analysis in the purely kinematical regime \citep{Bugli:2014}.

Similar GRMHD simulations have been recently employed to model the dynamics and emission of the central core of the galaxy M87 \citep{EHTCollaboration:2019}, comparing with observational data including the very first image of the \emph{shadow} of a black hole's event horizon \citep{EHTCollaboration:2019a}. Contrary to the standard numerical initial setup, where a subdminant but non-negligible magnetic field ($p_\mathrm{mag}/p \sim 10^{-2}$) is present in the disk right from the start \citep[e.g.][]{Porth:2019}, here we initialize the simulation with an extremely small magnetic field ($p_\mathrm{mag}/p \sim 10^{-9}$), which is later self-consistently amplified during the evolution by the $\alpha - \Omega$ dynamo process.

A linear kinematic exponential growth, followed by an other one with reduced rate when accretion becomes important, is observed for a variety of dynamo parameters. Both rates increase linearly with the dynamo parameter at least for the limited range explored here. The dynamo is clearly the main driver, with the second, reduced stage due to the fact that due to accretion the density in the disk is modified and the non-dimensional dynamo number $C_\xi$ reduces in time, especially for runs with a stronger dynamo. At later times the accretion process and the stronger field are capable of affecting the overall structure of the disk itself and the growth of the magnetic field ceases, reaching a saturation phase where the magnetic field is approximately constant in time. The presence of an explicit quenching term in the $\alpha$-dynamo shortens the linear phase but seems not to affect the final saturation stage, occurring roughly for similar values of the magnetic field strength (within a factor of three). The quenching also helps in avoiding the formation of a few pathological structures with highly magnetized vortices evacuating the plasma in the outer disk regions, and the overall dynamics is more regular.

In spite of the widely recognized importance of MRI for global simulation of disks around black holes \citep[e.g.][and references therein]{Bugli:2018}, and in spite of our simulations having the sufficient spatial accuracy to resolve such instability (the MRI \emph{quality factor} $Q$ exceeds 10 after a few rotational periods), this does not seem to play a major role here, if not as an initial trigger for the turbulent cascade and accretion onto the black hole. We have tested this hypothesis by switching off the dynamo term at the end of each growth phase (in different runs with otherwise the same parameters): the field starts to decrease immediately, and MRI seems to be just capable of balancing dissipation at small scales, reaching a steady (turbulent) state at late times.

By assuming the approximation of an optically thin plasma, as expected for Sgr A*, the accreting supermassive black hole of our Galaxy, we have computed on top of our simulations the expected emission, at millimiter wavelengths, for such source. A two-temperature plasma and all the recipes commonly employed for ADAF/RIAF systems have been used, obtaining very reasonable results, compared to previous works \citep{Moscibrodzka:2009,Moscibrodzka:2014}. This confirms that the dynamo action, believed to occur in these systems due to small-scale turbulence, is capable of amplifying the magnetic fields, in a self-consistent way, up to the values required to reproduce the observations.

All simulations have been performed with our \texttt{ECHO} code \citep{DelZanna:2007}, which has recently successfully participated to a code comparison project by the EHT collaboration \citep{Porth:2019}, in the upgraded version to include non-ideal resistive and dynamo effects in the Ohm's law \citep{Bucciantini:2013,DelZanna:2018}. From a computational point of view, we have here improved the implicit version of the conservative-to-primitive inversion step, by providing for the first time the analytical Jacobian matrix needed in the 3-D Newton-Raphson scheme. A similar approach has been recently employed for purely resistive schemes \citep[][]{Mignone:2019,Ripperda:2019}.

To conclude, we believe that non-ideal resistive-dynamo models of accreting disks around black holes represent a necessary upgrade to the existing ideal ones. However, for a detailed comparison against the revolutionary images by the EHT collaboration, fully 3-D simulations and ray-tracing techniques in curved spacetimes are certainly required.

\section*{Acknowledgements}

The authors are grateful to the anonymous referee for very helpful suggestions. Numerical calculations have been made possible through a CINECA-ISCRA project and a CINECA-INFN agreement, providing access to resources on MARCONI at CINECA. We acknowledge financial support from the ‘Accordo Attuativo ASI-INAF n. 2017-14-H.0 Progetto: on the escape of cosmic rays and their impact on the background plasma’ and from the INFN Teongrav collaboration. MB acknowledges support from the European Research Council (ERC starting grant no. 7153 68 -- MagBURST).


\appendix
\section{The coefficients for the electric field and for its Jacobian}

In the present Appendix we show how Eq.~\eqref{eq:implicit_final} has been derived and we provide the expressions for the required coefficients of $\Gamma$ and of their derivatives, to be used in Eq.~\eqref{eq:E_jacobian}. Since we are working with spatial vectors alone, involving the 3-D metric tensor $\gamma_{ij}$ (not diagonal in Kerr-Schild coordinates), we use here for simplicity the standard vector notation, for which $\bm{v}$ is employed rather than $v^i$.

The implicit step of the IMEX scheme for the electric field can be written as
\be
\begin{split}
 \bm{E} =  \bm{E}_\star - & \tilde{\eta}^{-1} \Gamma \{ \bm{E} 
+ \bm{v} \times \bm{B} - (\bm{E} \cdot \bm{v})\bm{v} \\
& - \xi [ \bm{B} - \bm{v} \times \bm{E} - (\bm{B} \cdot \bm{v}) \bm{v}  ] \}
\end{split}
\ee
and after the introduction of $\tilde{\bm{u}}=\Gamma\bm{v}$, for which $\Gamma^2 = 1 + \tilde{u}^2$, we ca rewrite the above expression as
\be
\begin{split}
( \Gamma +  \tilde{\eta}) \bm{E} & =  \tilde{\eta}\bm{E}_\star 
- \tilde{\bm{u}} \times \bm{B} + (\bm{E} \cdot \tilde{\bm{u}}  )\tilde{\bm{u}}/\Gamma  \\
& + \xi \Gamma\bm{B}  
- \xi ( \tilde{\bm{u}}\times \bm{E})
- \xi (\bm{B} \cdot \tilde{\bm{u}}) \tilde{\bm{u}}/\Gamma .
\end{split}
\label{eq:E}
\ee
The dot product with $\tilde{\bm{u}}$ allows one to write
\be
(1 +  \tilde{\eta} \Gamma)  ( \bm{E} \cdot \tilde{\bm{u}}) = 
\tilde{\eta} \Gamma ( \bm{E}_\star \cdot \tilde{\bm{u}}) + \xi ( \bm{B} \cdot \tilde{\bm{u}} ) ,
\label{eq:dot}
\ee
whereas the curl with $\tilde{\bm{u}}$ (necessary when $\xi\neq 0$) leads to
\be
\begin{split}
& ( \Gamma +  \tilde{\eta}) ( \tilde{\bm{u}}\times \bm{E}) = \tilde{\eta} ( \tilde{\bm{u}} \times \bm{E}_\star) 
+ (\Gamma^2-1)  \bm{B}    \\
&  - (\bm{B} \cdot \tilde{\bm{u}}) \tilde{\bm{u}} + \xi \Gamma ( \tilde{\bm{u}} \times \bm{B}) 
+ \xi [(\Gamma^2-1) \bm{E} - (\bm{E} \cdot \tilde{\bm{u}}) \tilde{\bm{u}}  ] .
\end{split}
\label{eq:curl}
\ee
Eqs.~(\ref{eq:dot} - \ref{eq:curl}) can be plugged into Eq.~(\ref{eq:E}) to derive an explicit expression for $\bm{E}$. In a compact form the required expression is
\be
\begin{split}
& A_0 \, \bm{E} = \tilde{\eta}  \bm{E}_\star  
+ A_1 \, ( \bm{E}_\star \cdot \tilde{\bm{u}}) \tilde{\bm{u}}
+ A_2  \, \tilde{\bm{u}} \times \bm{E}_\star \\
& + A_3 \, \bm{B} 
+ A_4 \, ( \bm{B} \cdot \tilde{\bm{u}} ) \tilde{\bm{u}}
+ A_5  \, \tilde{\bm{u}} \times \bm{B} ,
\end{split}
\ee
that is precisely Eq.~\eqref{eq:implicit_final}, where we have defined six new coefficiente, functions of $\Gamma$ alone, namely
\begin{eqnarray}
A_0(\Gamma) & = & \Gamma +  \tilde{\eta}+ \xi^2 \frac{\Gamma^2-1}{ \Gamma +  \tilde{\eta} }, \\
A_1(\Gamma) & = & \frac{\tilde{\eta}}{1 +  \tilde{\eta} \Gamma} 
+ \xi^2 \frac{\tilde{\eta}\Gamma}{ (\Gamma +  \tilde{\eta} )(1 +  \tilde{\eta} \Gamma ) }, \\
A_2(\Gamma) & = & - \xi \frac{\tilde{\eta}}{\Gamma +  \tilde{\eta}} , \\
A_3(\Gamma) & = & \xi \frac{1 + \tilde{\eta}\Gamma }{ \Gamma +  \tilde{\eta} }, \\
A_4(\Gamma) & = & \xi \, \frac{1 -  \tilde{\eta}^2 + \xi^2}{(\Gamma +\tilde{\eta} )(1+\tilde{\eta} \Gamma )}, \\
A_5(\Gamma) & = & -1 - \xi^2 \frac{\Gamma}{\Gamma +  \tilde{\eta}} .
\end{eqnarray}
The derivatives of the above coefficients, to be plugged into Eq.~\eqref{eq:E_jacobian}, are
\begin{eqnarray}
\dot{A}_0(\Gamma) & = & 1+ \xi^2 \frac{1 + \Gamma^2 + 2\tilde{\eta} \Gamma }{ (\Gamma +  \tilde{\eta})^2 }, \\
\dot{A}_1(\Gamma) & = & - \frac{\tilde{\eta}^2}{(1 +  \tilde{\eta} \Gamma)^2} 
- \xi^2 \frac{\tilde{\eta}^2( \Gamma^2 -1) }{(1 +  \tilde{\eta} \Gamma)^2(\Gamma +  \tilde{\eta})^2}  , \\
\dot{A}_2(\Gamma) & = & \xi \frac{\tilde{\eta}}{(\Gamma +  \tilde{\eta})^2} , \\
\dot{A}_3(\Gamma) & = & \xi \frac{\tilde{\eta}^2 - 1}{(\Gamma +  \tilde{\eta})^2} , \\
\dot{A}_4(\Gamma) & = & - \xi \, \frac{ (1 -  \tilde{\eta}^2 + \xi^2) (1 + \tilde{\eta}^2 + 2\tilde{\eta}\Gamma)  }{(\Gamma +  \tilde{\eta} )^2(1 +  \tilde{\eta} \Gamma )^2}, \\
\dot{A}_5(\Gamma) & = & -  \xi^2 \frac{\tilde{\eta}}{(\Gamma +  \tilde{\eta})^2} ,
\end{eqnarray}
and thanks to the above expressions the Jacobian of the electric field can be computed.

\end{document}